# Title:
# Efficient Energy Transport in an Organic Semiconductor Mediated by Transient Exciton Delocalization

**Authors**: Alexander J. Sneyd[1,+], Tomoya Fukui[2,3,†,+], David Paleček[1], Suryoday Prodhan[4], Isabella Wagner[5], Yifan Zhang[2,3], Jooyoung Sung[1], Sean M. Collins[6], Thomas J. A. Slater[7], Zahra Andaji-Garmaroudi[1], Liam R. MacFarlane[2,3], J. Diego Garcia-Hernandez[2,3], Linjun Wang[8], George R. Whittell[3], Justin M. Hodgkiss[5], Kai Chen[5,9,10], David Beljonne[4,*], Ian Manners[2,3,*], Richard H. Friend[1], and Akshay Rao[1,*]

**Affiliations:**
[1]Department of Physics, Cavendish Laboratory, University of Cambridge, Cambridge CB3 0HE, United Kingdom
[2]Department of Chemistry, University of Victoria, Victoria, BC V8P 5C2, Canada
[3]School of Chemistry, University of Bristol, Bristol BS8 1TS, United Kingdom
[4]Laboratory for Chemistry of Novel Materials, University of Mons, Mons 7000, Belgium
[5]MacDiarmid Institute for Advanced Materials and Nanotechnology and School of Chemical and Physical Sciences, Victoria University of Wellington, Wellington 6010, New Zealand
[6]School of Chemical and Process Engineering and School of Chemistry, University of Leeds, Leeds LS2 9JT, UK
[7]Electron Physical Sciences Imaging Centre, Diamond Light Source Ltd., Oxfordshire OX11 0DE, UK
[8]Center for Chemistry of Novel & High-Performance Materials, and Department of Chemistry, Zhejiang University, Hangzhou 310027, China
[9]Robinson Research Institute, Faculty of Engineering, Victoria University of Wellington, Wellington 6012, New Zealand
[10]The Dodd-Walls Centre for Photonic and Quantum Technologies, Dunedin 9016, New Zealand
*Correspondence to: ar525@cam.ac.uk, imanners@uvic.ca, david.beljonne@umons.ac.be..
[+]These authors contributed equally
[†]Present address: Tokyo Institute of Technology, R1-1, 4259 Nagatsuta, Midori-ku, Yokohama, Kanagawa, 226-8503, Japan

**Abstract**: Efficient energy transport is highly desirable for organic semiconductor (OSC) devices such as photovoltaics, photodetectors, and photocatalytic systems. However, photo-generated excitons in OSC films mostly occupy highly localized states over their lifetime. Energy transport is hence thought to be mainly mediated by the site-to-site hopping of localized excitons, limiting exciton diffusion coefficients to below ~$10^{-2}$ cm$^2$/s with corresponding diffusion lengths below ~50 nm. Here, using ultrafast optical microscopy combined with non-adiabatic molecular dynamics simulations, we present evidence for a new highly-efficient energy transport regime: transient exciton delocalization, where energy exchange with vibrational modes allows excitons to temporarily re-access spatially extended states under equilibrium conditions. In films of highly-ordered poly(3-hexylthiophene) nanofibers, prepared using living crystallization-driven self-assembly, we show that this enables exciton diffusion constants up to $1.1 \pm 0.1$ cm$^2$/s and diffusion lengths of $300 \pm 50$ nm. Our results reveal the dynamic interplay between localized and delocalized exciton configurations at equilibrium conditions, calling for a re-evaluation of the basic picture of exciton dynamics. This establishes new design rules to engineer efficient energy transport in OSC films, which will enable new devices architectures not based on restrictive bulk heterojunctions.

**Teaser:** Precisely tuning an organic semiconductor's crystallinity enables excitons to move 2-3 orders of magnitude faster than previously thought possible.

**MAIN TEXT**

The efficient transport of energy in the form of excitons is crucial to the functioning of light-harvesting devices based on organic semiconductors (OSCs), as well as naturally-occurring light-harvesting complexes (LHCs). In the former, it allows excitons generated in the material's bulk to reach charge-generating heterojunctions, and in the latter, it allows energy to reach reaction centres. However, strong exciton-phonon couplings and electron-hole Coulomb interactions cause the localization of exciton wavefunctions in OSCs, and current models consider the physics of OSCs to be dominated by these localized states. Energy transport in particular is generally considered through the framework of Förster resonance energy transfer (FRET), where localized excitons hop incoherently from site to site (*1–3*). This mechanism naturally limits the distance excitons can travel (*3*). In LHCs by contrast, intriguing experimental and theoretical evidence has been put forward in recent years suggesting that coherent energy transport via short-lived superpositions of exciton states might contribute to the function of LHCs (*4, 5*). Within this framework, electronic and/or vibronic coherences (*6*) enable delocalized excitons to move tens of nanometres on the timescale of hundreds of femtoseconds following photoexcitation, before localization occurs. In the case of LHCs, such timescales are sufficient to allow for efficient energy transfer from antennae complexes to reaction centres. But for devices based on OSC films such as photovoltaics, photodetectors and photocatalytic systems, the ideal length scales for exciton transport would be matched to the material's absorption depth – typically 100-200 nm, and the coherent transport of energy over such distances is extremely challenging. Hence, energy transport in OSCs has remained largely restricted to FRET, and despite many decades of work, singlet exciton diffusion lengths ($L_D$) in OSC films are typically limited to about 10 nm, with associated diffusion constants ($D$) on the order of $10^{-3}$-$10^{-4}$ cm$^2$/s (*3, 7*).

The restriction to FRET/localized excitons in OSCs represents a significant obstacle for devices, as the short $L_D$'s lower device efficiencies and necessitate the use of nanoscale bulk heterojunctions which compromise other properties such as charge extraction and stability. Improved energy transport has thus remained a long-outstanding goal. Promisingly, several recent observations have shown that in isolated self-assembled nanostructures, exciton diffusion lengths and diffusion constants can be considerably higher (*8–10*). For example, oligomeric polyfluorene nanofibers can show $L_D$'s of 270 nm (*9*) and tubular porphyrin aggregates display $D$'s of 3-6 cm$^2$/s (*10*). However, such behaviour has not been extended to device-relevant films, nor to materials with broad visible-light absorption. Most critically of all, the fundamental question of how energetic and structural order can allow such a large increase in exciton mobility remains unanswered.

Herein we study films of well-ordered poly(3-hexylthiophene) nanofibers which absorb broadly across the visible spectral region and provide experimental and theoretical evidence for a new highly-efficient energy transport regime in OSCs: *transient delocalization*. In this scheme, excitons still spend most of their lifetime in localized states as in FRET, but *retain access* to delocalized states by exchanging energy with vibrational modes under *equilibrium conditions* i.e. well after the initial (short-lived) period of delocalization following photoexcitation. We show that this allows for the efficient transport of excitons over long distances at room temperature; in films of the nanofibers we measure $D$ to be $1.1 \pm 0.1$ cm$^2$/s via ultrafast optical microscopy and spectroscopy, giving an estimated $L_D$ of $300 \pm 50$ nm, which is higher than the absorption depth of the material.

## Results

**Model organic semiconductor system**

Fig. 1 displays the explored model system – regioregular poly(3-hexylthiophene) (rr-P3HT) nanofibers (NFs). To create the P3HT NFs, we employ living crystallization-driven self-assembly (CDSA) for its ability to produce well-ordered, colloidally-stable, and uniform nanostructures via the epitaxial crystallization of polymer-based amphiphiles – including those based on conjugated polymers (*11–16*). Importantly, this seeded-growth technique also allows for exquisite size control, the formation of segmented assemblies from different building blocks, and surface functionalization, so its control and versatility makes it attractive for device fabrication (*9, 17–20*). P3HT was chosen as the base material since it strongly absorbs visible light and has been widely studied for several decades as a model system for harvesting energy in organic photovoltaics, photodetectors, and photocatalysts. Using a recently described procedure, molecularly dissolved phosphonium borate-terminated P3HT amphiphile P3HT$_{30}$-[PPh$_3$Me][BPh$_4$] in THF was added to short seeds of the same material (see Fig. 1 and Methods). This yielded a sample of uniform fiber-like micelles (referred to as *nanofibers*; NFs) with an average length of 890 nm (length dispersity = 1.08). The NFs have a height of 4.5 nm as determined by AFM (fig S4), a width of 12.8 nm as determined by TEM, and possess a solvated corona corresponding to the phosphonium group at the terminus of polythiophene, which permits colloidal stability due to the introduction of electrostatic repulsion between the nanofibers and additional opportunities for solvation (*21*). The predominant NF component is a crystalline core that consists of the regioregular poly(3-hexylthiophene) (rr-P3HT) segments (Fig. 1B). This crystalline rr-P3HT core has a well-defined crystal structure (as determined from X-ray diffraction measurements – see fig. S6), where the rr-P3HT chains pack side-by-side to afford good π-π orbital overlap, and hence strong electronic interactions between neighbouring chains. The NFs appear to be exclusively crystalline, as demonstrated by the homogenous profile observed both in TEM (Fig. 1C and S5) and confocal laser scanning microscopy (fig. S14), as well as the *lack* of photoluminescence that would normally be attributed to amorphous chains (see S23 for more details).

We prepare films from the NFs (see Methods) to produce dense meshes of NFs. To characterize these films, we use low-dose scanning electron diffraction (SED), which gives Bragg diffraction spots corresponding to the π-π stacking of the NFs with ~4 nm resolution – see Fig. 1C. By analysing the angle of the diffraction spots, an 'orientation' map can be formed (Fig. 1D), where the colour records the angle of the pair of diffraction spots, i.e. the orientation of the π-π stacking planes. While the NFs are mostly orientated randomly with respect to one another (as indicated by the different overall colours of each NF), the NFs on their own tend to exhibit large straight sections with a consistent colour, indicating the rr-P3HT chains are fully aligned over these domains. To parametrize the length of these straight, aligned domains, we introduce the *persistence length*, $P$

(see S.I. for more details and fig. S9 for an illustration). The cumulative distribution of the extracted $P$'s in Fig. 1E shows that the NFs tend to exhibit large persistence lengths, with a mean value of $\langle P \rangle = 80$ nm and detected lengths up to $P > 300$ nm. This establishes excellent long-range ordering of rr-P3HT chains in the NFs when in the films; a consequence of the epitaxial growth mechanism inherent to living CDSA. The measurement of $\langle P \rangle$ is limited in this case by out-of-plane bending of the NFs in the films, which introduces deviations from the Bragg condition.

The NFs' long-range structural order contrasts with that which is typically observed in polymer films, including self-organizing high-molecular-weight spincoated rr-P3HT (herein referred to as "rr-P3HT"), which has been studied extensively for over two decades. Although both the NF films and rr-P3HT exhibit the same π-π stacking motif (*22*) – exemplified in Fig. 2C by the similar absorption profile – Fig. 1F illustrates how rr-P3HT consists of both amorphous and crystalline regions, with a high degree of energetic and structural disorder, and limited long-range ordering of polymer chains (*22, 23*).

**Direct measurement of efficient energy transport**

We use transient-absorption microscopy (TAM) to directly image the interchain exciton diffusion along the NFs in the NF films. As shown in Fig. 2A-B, this technique employs wide-field imaging via a time delayed ~100 fs probe pulse, following excitation with a diffraction-limited ~30 fs pump pulse centred at 540 nm - see refs (*24, 25*) for further information. We measure a differential transmission distribution ($\Delta T/T$) at a probe wavelength of 600 nm as a function of pump-probe delay – see S.I. and Fig. 2 for further details. Since the $\Delta T/T$ distribution is proportional to the photo-generated exciton distribution, $n(t,x,y)$, expansion of the $\Delta T/T$ distribution (see inset in Fig. 2A) is related to the movement of excitons. This movement can be parametrized by the mean square displacement, $MSD(t)$, of the ensemble, which is defined by: $MSD(t) \equiv \sigma^2(t) - \sigma^2(0)$, where $\sigma^2(t)$ is the spatial variance of the exciton distribution extracted by fitting a 2D Gaussian to each $\Delta T/T$ image.

Figure 2D plots the $MSD(t)$ for a representative measurement. We note that at early times we observe signatures of exciton-exciton annihilation (EEA) at the fluence used here (4 µJ/cm²) which would result in an overestimation of $D$ at early times. However, for time delays beyond 10 ps the extracted $D$ values are not affected by pump fluence (see fig S27 and detailed discussion in S.I.). It can be seen in Fig. 2D how $MSD(t)$ is linearly related to time in this region, allowing for a reliable extraction of $D$ ($MSD(t) = 2Dt$, see S.I. for details). Performing many such individual measurements across different locations on different films allows us to build up statistics for $D$, as shown in Fig. 2E. Taken together, the average diffusion constant is $D_{NF} = 1.1 \pm 0.1$ cm²/s.

To rule out systematic artefacts, we also performed measurements on two control samples: rr-P3HT films, and films of chloroform (CHCl₃)-treated NFs (where application of CHCl₃ on the film disrupts the long-range order of the nanofibers – see Methods and fig. S12 for details). The resultant diffusion constants were both below the experimental detection limit of about 0.1 cm²/s. For rr-P3HT this is expected; $D$ is known to be significantly below 0.1 cm²/s for solution-processed films (*26, 27*). Furthermore, a low value of $D_{NF+CHCl_3}$ is also to be expected if indeed $D_{NF}$ is high because of the NF's intrinsic order. We are therefore able to rule out any systematic artefacts contributing to $D_{NF}$.

We also analyzed EEA effects in femtosecond transient-absorption spectroscopy (fs-TA) and time-resolved photoluminescence (TRPL) to estimate $D_{NF}$ – see S.I. for full details. In these measurements, we (indirectly) find that $D$ ranges from 0.2 to 0.8 cm²/s for fs-TA and 0.2 to 1.0 cm²/s for TRPL, which broadly agree with the more accurate $D_{NF}$ value directly found using TAM.

Our experimental data therefore establishes a diffusion constant of $D_{NF} = 1.1 \pm 0.1$ cm²/s in the NF films. This value is remarkable considering that previous reports of exciton diffusion within crystalline regions of rr-P3HT gave $D$ as $2 \times 10^{-2}$ cm²/s (*28*), $7.9 \times 10^{-3}$ (*27*), and $1 \times 10^{-2}$ cm²/s (*26*), and most reported values for OSC films are in the range of $10^{-3}$-$10^{-4}$ cm²/s (*3, 27, 29*). Significantly, this $D_{NF}$ value is found for a OSC *film* – which are most relevant for devices – as opposed to previous observations of efficient (equilibrium) transport which were found for isolated and pristine single nanostructures (*8–10*). Combined with the exciton lifetime of $\tau = 400 \pm 100$ ps extracted using fs-TA (see fig. S19), we estimate a corresponding diffusion length of $L_{NF} = \sqrt{2D\tau} = 300 \pm 50$ nm. This calculation of $L_{NF}$ assumes that $D$ remains constant over the exciton's lifetime, which is a reasonable assumption as we detail later. This value is an order of magnitude higher than previously reported values for rr-P3HT of 20 nm and 27 nm (*26, 27*), or indeed any solution-processed conjugated polymer film.

**Energetic disorder**

In any mechanistic picture of exciton transport, energetic disorder due to local imperfections or variations in the crystalline packing of molecules or polymer chains is expected to restrict exciton diffusion since excitons may get trapped at low-energy sites (*2, 3, 7, 30–33*). We hence characterise the energetic order of the NF and rr-P3HT films. Using photothermal deflection spectroscopy (PDS), we extract an Urbach energy ($E_u$) – the width of the sub-band gap tail states – of $E_{u;NF} = 29 \pm 1$ meV for the NF films (see Fig. 3A). This value is significantly lower than the value we find for rr-P3HT ($53 \pm 6$ meV), in agreement with literature (*34*). To our knowledge, $E_{u;NF}$ is the lowest value measured for a P3HT-based system, and among the lowest reported for any OSC (*35–37*). We also note that the NF's baseline absorption at low energies (<1.7 eV) is over 40 times lower than that of rr-P3HT, indicating a greatly reduced density of deep trap states. This lack of deep traps in the NFs is understandable given the living CDSA growth mechanism, since the slow epitaxial growth from the seed micelles will tend to exclude chains with defects, prevent twists or kinks in the chains, and overall result in much purer crystal than spincoating.

To examine the effect of this energetic disorder on the relaxation of excitons within the excitonic density of states (EDOS) (which occurs via exciton diffusion to lower-energy sites) we use a combination fs-TA and TRPL. As shown in Fig. 3B, in the NF films, the position of the $E_{0-1}(t)$ vibronic peak relaxes by only ~20 meV over 1 ps following photoexcitation. In contrast, excitons in rr-P3HT rapidly relax ~80 meV by 1 ps. The fact that this value (20 meV) and the Urbach energy (29 meV) are comparable to $k_B T$ at room temperature (26 meV) suggests that thermal fluctuations can easily overcome energetic disorder within the NF films and that excitons do not become deeply trapped. We also highlight that the average $E_{0-1}$ position in TRPL changes only 6 meV between the time ranges 10-30 ps and 30-375 ps (see fig. S24), and so the $D_{NF}$ value extracted from 10-30 ps is expected to be valid at later times, lending validity to our estimate of $L_{NF}$ previously.

**Exciton transport mechanism**

We now examine the mechanistic nature of the exciton transport. One possibility is the classical FRET-based picture, where localized excitons hop from site to site. This is the predominant description given for exciton transport in OSC films. However, in our case, we find that a simple Förster hopping model compares very poorly with experiment, giving an upper value of $D_{NF} \sim 2 \times 10^{-3}$ cm²/s (a value which is similar to those given in previous reports; see SI for more details) (*28, 31, 38*). Alternatively, one might speculate that the NFs' behaviour is similar to that proposed in LHCs, where short-lived delocalized states at early times are mainly responsible for the exciton transport. Again, however, this picture does not fit the experimental data; we observe no anomalous transport at early times, and $D_{NF}$ is extracted from 10 ps onwards by which time excitons will have reached quasi-equilibrium conditions (as shown by the data in Fig. 3B)

To model the time-dependent behaviour of the exciton wavefunction and hence model exciton transport we employ non-adiabatic molecular dynamics simulations. The diffusion of excitons was simulated in one-dimensional stacks of ~300 P3HT chains as shown in Fig. 4B using a mixed quantum-classical, crossing-corrected variant of the subspace surface hopping algorithm (*39–41*), which incorporates stochastic non-adiabatic transitions between different adiabatic potential energy surfaces (*42*). The inclusion of long-range exciton couplings in the simulation is physically realistic given the mean persistence length of $\langle P \rangle = 80$ nm as measured by SED. The time-evolved exciton wavefunctions are calculated from 10,000 individual trajectories, and are then averaged to compute an ensemble exciton density and $MSD(t)$ - see Fig. 4B-C. The corresponding diffusion constant of $D_{sim} = 0.22$ cm²/s is the same order of magnitude as our experimental value of $D_{NF} = 1.1 \pm 0.1$ cm²/s.

To interrogate the reason for such a high $D_{sim}$ value we assess the role of exciton delocalization. We compute the inverse participation ratio (IPR) – a proxy for the interchain exciton delocalization – and find a mean IPR of ~1 (see S.I.), demonstrating that excitons are in fact *mostly* localized on a single chain. However, analysis of the time-dependent IPR profile of the excitonic wavefunction over a single, randomly selected, trajectory (see Fig. 4D-E) reveals something curious: while the excitation remains mostly confined over a single chain at the bottom of the EDOS, it occasionally momentarily reaches slightly higher-energy adiabatic states that spread over several neighbouring chains. This transient access to delocalized states occurs thanks to the continuous energy exchange with vibrational modes in the material, serving to kick the exciton over large distances on a short time span. As shown schematically in Fig. 4F, the exciton sporadically 'surfs' on the EDOS. We propose that this transient delocalization enables the large values of $D$ experimentally measured here. We note that the proposed model is similar to the mobility edge model of charge transport (*43*) in OSCs, and a similar scenario has been shown to accurately reproduce *charge* mobilities in organic crystals (*44*).

Interestingly, this 'transient delocalization' scenario is critically dependent on the presence of long-range exciton couplings; as shown in Fig. 4B-C, when all non-nearest neighbour interactions are set to zero (thus adopting the tight-binding approximation used for charges), there is no transient access to delocalized states, which reduces $D$ by three orders of magnitude to $D = 6 \times 10^{-4}$ cm$^2$/s, which is similar to the value for $D$ obtained in rr-P3HT films previously (*3, 26, 27, 29, 45*). We note that stacking faults, point defects or grain boundaries would shorten the effective range of the excitonic interchain interactions, thereby deactivating transient delocalization. Crucially, this is avoided in our case, with multiple techniques demonstrating high crystallinity in the NF films, and the long persistence length of 80 nm observed in SED confirming the existence of long-range excitonic interchain interactions. This highlights the advantage of our highly-controlled living CDSA synthetic process in enabling long-range structural order and hence transient delocalization.

On the other hand, $D$ is only moderately sensitive to energetic disorder; when the inhomogeneous broadening is halved or doubled, the extracted $D$ only increases or decreases by ~40% respectively (see fig. S29), suggesting that long-range interactions smooth out, to some extent, energetic inhomogeneity in the material (*35*). A key point may be the absence of deep energetic traps, as indicated by the PDS data, in comparison to conventional materials. Once trapped in such sites the wavefunction would not be able to access the delocalized states via energy exchange with vibrational modes.

**Discussion**

In summary, we have demonstrated that well-ordered visible-light-absorbing films of P3HT NFs prepared via living CDSA exhibit unprecedented exciton diffusion properties, and to explain these results, propose a new energy transport framework: transient delocalization. In this new scheme, the lowest-energy exciton state is still highly localized (as is the case for most OSCs), however energy exchange with vibrational modes can temporarily promote excitons to delocalized states, with this transient access responsible for a 2-3 order-of-magnitude enhancement in the energy transport rate. This new picture challenges the widely held notion that excitons in OSCs are localized under equilibrium conditions; instead, they can experience large fluctuations in their spatial extent given the right structural and energetic conditions, with briefly-occupied delocalized states playing an unexpectedly large role in the transport process. Importantly, this highly-efficient regime of energy transport should be achievable across a wide range of device-relevant organic semiconductors as is the case for the equivalent charge transport mechanism (transient charge delocalization/localization). We expect further exploration of the emergent transient delocalization behaviour to prove highly fruitful, with particularly interesting directions including its temperature dependence, its dependence on the unimer's length, and the inclusion of non-local electron-phonon couplings into the model. This may help reveal further design rules in addition to our two foundational rules that energetic disorder be comparable to $k_B T$ so that energy exchange with vibrational modes may provide access to extended states, and long-range structural order over 10's of nanometres so that these extended states can be supported. Ultimately, engineering transient delocalization in devices may thus provide a route towards improved efficiencies in a range of organic semiconductor devices, particularly since the crystalline order that transient delocalization demands would also be beneficial for properties such as charge extraction.

Our work also demonstrates the power of living CDSA for engineering efficient transport in visible-light-absorbing thin films. This may help improve the performance of conventional devices, and interestingly, opens up new device architectures for photovoltaics, photodetectors, photocatalytic systems and photon up/down-conversion systems that would no longer be limited by the nanoscale bulk heterojunction architecture and could funnel energy over hundreds of nanometres to active sites/heterojunctions. For example, recent advances in living CDSA have demonstrated the ability to grow nanostructures directly off silicon surfaces (*19*). This could allow for semiconducting inorganic-organic heterojunctions where high exciton diffusivity in conjunction with strong visible-light absorption could be exploited for enhanced photo-detection.

**Materials and Methods**

Preparation of P3HT nanofibers via living CDSA for the preparation of films

The solution of the seed micelles (Conc. = 0.5 mg/mL) in benzonitrile/THF mixture (benzonitrile:THF = 8:2 v/v) was warmed at 35 ºC for 5 min. Appropriate amounts of unimers in THF were added to the seed micelles. After the mechanical shaking for 10 sec, the solutions were aged for 12 h. THF and some benzonitrile was evaporated by a gentle N$_2$ gas flow for 3 h (final Conc. = 1.0 mg/mL). TEM analysis was used to evaluate the length of the resulting micelles after dilution to 0.02 mg/mL using benzonitrile.

Preparation of films for spectroscopy

All film preparation was conducted in an N$_2$ atmosphere. Unless otherwise stated, the samples were deposited on 22 mm × 22 mm No. 1.5 borosilicate 170 $\mu$m thick glass coverslips (Deckgläser) which were pre-cleaned via 10 mins sonication in Acetone/water = 8:2 v/v and then isopropanol. The films were subsequently encapsulated in the same N$_2$ glovebox by placing a second, smaller coverslip atop the sample substrate with a ~200 $\mu$m thick carbon tape spacer and sealing with epoxy resin. All spectroscopy experiments were performed at room temperature.

The NF films used for spectroscopy were prepared via drop-casting to produce films consisting of a dense, randomly-orientated mesh of NFs stacked atop each other. Each film was allowed 12 hrs for the benzonitrile solvent to fully dry. These NF films were all prepared from the same original batch of NFs. The rr-P3HT films were made via spin-coating solutions of rr-P3HT (96% regioregularity, average molecular weight of 90 kDa, polydispersity index of 2.3) in chlorobenzene at 2000 rpm. The chloroform-treated NF films were prepared by first dissolving the NFs in $CHCl_3$. The resultant unimer solution was drop-cast onto a glass slide heated to ~80°C to rapidly remove the solvent, thereby preventing any substantial long-range crystallization. $CHCl_3$ was then reapplied to the film to dissolve the unimers a second time so that the solvent was now exclusively $CHCl_3$, and was then rapidly dried again. All the films were optimized for spectroscopy so that their absorption peaks (which all match up very well – see S.I.) corresponded to ~0.3 OD in the visible region.

Scanning electron diffraction

Samples were prepared by drop-casting the nanofibers onto carbon-coated copper grids (Agar Scientific). Scanning electron diffraction (SED) data was acquired using a JEOL ARM300CF fitted with an ultrahigh-resolution pole piece, a cold field emission gun, and aberration correctors in both the probe-forming and image-forming optics (Diamond Light Source, UK). The instrument was operated at 300 kV. A nanobeam configuration was obtained by switching off the aberration corrector in the probe-forming optics and using a 10 μm condenser aperture to obtain a convergence semiangle <1 mrad and a diffraction-limited probe diameter of ca. 3 nm. The probe current was measured using a Faraday cup as ca. 2 pA, and the exposure time was 1 ms per probe position. The estimated electron fluence, assuming a disk-like probe, was <20 $e^-$ $Å^{-2}$. A diffraction pattern was acquired at every probe position using a quad-chip Merlin-Medipix hybrid counting-type direct electron detector (Quantum Detectors, UK) with 512×512 pixels. SED was obtained in a "blind scanning" point-and-shoot workflow, to minimize the total electron fluence the specimen received.

SED data were processed using pyxem-0.11.0 (*46*). The diffraction patterns were aligned and calibrated following previously reported procedures (*47*). Briefly, the calibration of the scan step size and the diffraction pattern pixel size was performed using a standard 500 nm gold diffraction grating replica with latex spheres (Ted Pella). The cross-grating data was also used to determine residual elliptical distortions of the diffraction patterns due to the post-specimen optics. The rotation between the diffraction pattern and the real space orientation was calibrated using an $MoO_3$ standard (Agar Scientific). Processing of the diffraction patterns included: 1) centring the direct beam in each diffraction pattern within the data array using a cross-correlation routine, 2) applying an affine transformation to correct for elliptical distortion and rotation between the real space scan and the diffraction pattern, as determined by calibration.

Photothermal deflection spectroscopy

PDS sensitively measures absorption directly by probing the heating effect in samples upon absorption of light. Films were coated on a Spectrosil fused silica substrate and were immersed in an inert liquid FC-72 Fluorinert. They were then excited with a modulated monochromated light beam perpendicular to the plane of the sample.  A combination of a Light Support MKII 100 W Xenon arc source and a CVI DK240 monochromator was used to produce a modulated monochromated light beam. The PDS measurements were acquired by monitoring the deflection of a fixed wavelength (670 nm) diode laser probe beam following absorption of each monochromatic pump wavelength.

Transient-absorption microscopy

A thorough description of the setup can be found in ref. 23, but in brief: a Yb:KGW-based amplified laser system (Pharos, Light Conversion) provided 200 fs, 30 µJ pulses at 1030 nm with a 200 kHz repetition rate. The output beam was split to seed two broadband white light continuum (WLC) stages. The probe WLC was generated in a 3 mm yttrium aluminium garnet (YAG) crystal and adjusted to cover a 580-950 nm range via a fused-silica prism-based spectral filter. The pump WLC was generated in a 3 mm sapphire crystal to extend the WLC to 500 nm, and short-passed filtered at 575 nm. A set of chirped mirrors (2x for pump) and pair of fused silica wedges was used to pre-compress the pulses to ~100 fs (probe) and ~30 fs (pump). Shorter probe pulses (~10 fs) can be achieved by reducing the bandwidth, however this ultrafast time resolution was not needed, and the priority here was instead to probe at 600 nm (which corresponds to a ground-state bleach in the samples analyzed). A motorized delay stage (Newport) was used to delay the probe with respect to the pump. A clean mode for the pump was achieved with a pinhole. The pump was collimated to completely fill the aperture of the objective lens (x100, with an effective numerical aperture of 1.1) to deliver a near diffraction-limited spot with a ~270 nm FWHM. The probe was counter-propagated through the sample with a relatively large focal spot (~15 µm). The pump and probe polarizations were set to be parallel at the sample in order to maximize S/N. The transmitted probe pulse was collected by the same objective lens, was filtered with a 600 nm bandpass filter (10 nm FWHM), and imaged onto an EMCCD camera (Rolera Thunder, QImaging). An automatic focus control loop based on total internal reflection of a reference continuous wave laser (405 nm) was used to stabilize the focus position via an objective piezo (NP140, Newport). Differential imaging was achieved by modulating the pump beam at 30 Hz by a mechanical chopper.

Transient-absorption spectroscopy

The transient absorption experiments were performed using a Yb:KGW laser system (Pharos, Light Conversion) to provide 15.2 W at 1030 nm with a 38 kHz repetition rate. The probe beam was generated by focusing a portion of the fundamental in a 4 mm YAG crystal to generate a WLC. The pump beam was generated by a non-collinear optical parametric amplified (NOPA) seeded by WLC from a 3 mm YAG crystal mixed with a third harmonic pump (HIRO, Light Conversion) in a barium borate crystal (37° cut, type I, 5° external angle). The NOPA output was centred at 540 nm, and then compressed down to ~10 fs pulses using a pair of chirped mirrors and pair of fused silica wedges. The pump was delayed using a computer-controlled Thorlabs translation stage. A sequence of probe pulses with and without a pump was generated using a chopper wheel on the pump beam. After the sample, the probe pulse was split with a 950 nm dichroic mirror (Thorlabs). The visible part was then imaged with a Silicon photodiode array camera (Stresing Entwicklunsbüro; visible monochromator 550 nm blazed grating). The near-infrared portion of the probe was seeded to an IR spectrograph (1200 nm blazed grating) and imaged on an InGaAs photodiode array camera (Sensors Unlimited). Offsets for the differing spectral response of the detectors was accounted for in the post-processing of data.

Transient-gating photoluminescence spectroscopy

For the detection of the broadband PL on a sub-picosecond timescale, we used the transient grating PL technique (*48*). The output of a femtosecond Ytterbium Fiber Laser (Tangerine SP, Amplitude Systems) operating at 44 kHz pulses was split into pump and gate parts. The pump part was frequency-doubled to 515 nm using by an BBO crystal and focused to a 62 μm$^2$ spot to excite the samples. The PL was collimated and refocused onto the gate medium (2 mm undoped yttrium aluminium garnet crystal) using a pair of off-axis parabolic mirrors. For the gate part, ∼40 μJ 1030 nm output was split using a 50/50 beam splitter to generate the two gate beams and focused onto the gate medium at a crossing angle of approximately 8° and overlapped with the PL in a BOXCAR geometry. The two gate beams generate a laser-induced grating inside the gate medium, acting as an ultrafast gate to temporally resolve the broadband PL signals by diffracting the gated signal from the PL background. Two achromatic lenses collimated and focussed the gated signals onto the spectrometer entrance (Princeton Instruments SP 2150), and the gated PL spectra were measured by an intensified CCD camera (Princeton Instruments, PIMAX3). A long- and short-pass filter were used to remove the residual pump and intense 1030 nm gate, respectively. The time delay between pump and gate beams was controlled via a motorized optical delay line. For the transient PL spectrum at each time delay, 120,000 shots were accumulated. The setup can achieve detection bandwidth from 530 nm to 900 nm, limited by the colour filter pair, and instrument response function (∼230 fs).

Theory

We model the steady-state optical properties of 24-mer aggregates applying a Frenkel-Holstein model (solved in the two-particle approximation and with open boundary conditions). The Coulomb excitonic couplings are calculated using a transition density approach based on semi-empirical quantum-chemical calculations of the isolated polymer chains. Couplings of the electronic excitations to a high-frequency vibration at 0.18 eV (1450 cm$^{-1}$) and to low-frequency vibrations are explicitly included to reproduce the measured vibronic progression in optical absorption. The excitation energies are sampled from a non-correlated Gaussian-shape inhomogeneous disorder distribution. Under the (reasonable) assumption that steady-state PL occurs from the thermalized Excitonic Density of States (EDOS), i.e. in equilibrium conditions, simultaneously fitting the lineshape and linewidth in emission and the spectral Stokes shift yield the magnitude of the static (over the time scale of energy migration) and dynamic contributions to the energy disorder.

The same Hamiltonian is then employed to model the propagation of the excitons along the 1D stacks using a mixed quantum-classical, crossing-corrected variant of subspace surface hopping algorithm (*39–41*), which incorporates stochastic nonadiabatic transitions between different adiabatic potential energy surfaces (PES) (*42*). The dynamics of the full system is described by an ensemble of independent trajectories, where each trajectory occupies an 'active' PES at individual time steps (see details of the methodology in S.I.). Along each trajectory, the dynamics of the ions is treated classically by the Langevin equations of motion while the exciton wavefunction is propagated quantum-mechanically solving the time-dependent Schrödinger equation. The time-evolved exciton wavefunction is then utilized to calculate the MSD and the IPR, the latter providing a proxy for the intermolecular exciton delocalization. Linear evolution of MSD signifies an equilibrium diffusion regime is attained, out of which the diffusion coefficient (D) can be calculated. To study the non-adiabatic excited state dynamics, we consider a stack consisting of ~300 rr-P3HT chains (with open boundary conditions), initially exciting the central chain of the stack.

**Acknowledgments:** We thank C. Schnedermann for useful discussions.

**Funding:**
This project has received funding from the European Research Council (ERC) under the European Union's Horizon 2020 research and innovation programme (grant agreement No 758826).
This project has also received funding from the Engineering and Physical Sciences Research Council.
We thank the Winton Programme for the Physics of Sustainability and the Engineering and Physical Sciences Research Council for funding.
A.J.S acknowledges the Royal Society Te Apārangi; the Cambridge Commonwealth European and International Trust; and St John's College, University of Cambridge for their financial support.
I.M. thanks the Canadian Government for a Canada 150 Research Chair, NSERC for a Discovery Grant, and the University of Victoria for start-up funds.
T.F. thanks the Japan Society for the Promotion of Science (JSPS) for a postdoctoral fellowship for research abroad which was held in Bristol and Victoria.
J.D.G.H. thanks the Consejo Nacional de Ciencia y Tecnología (CONACYT) of Mexico for the provision of a scholarship. The authors thank the Centre for Advanced Materials and Related Technologies (CAMTEC) for the use of common facilities.

Computational resources in Mons were provided by the Consortium des Équipements de Calcul Intensif (CÉCI), funded by the Fonds de la Recherche Scientifique de Belgique (F.R.S.-FNRS) under Grant No. 2.5020.11, as well as the Tier-1 supercomputer of the Fédération Wallonie-Bruxelles, infrastructure funded by the Walloon Region under Grant Agreement No. 1117545.

S.P. is a Postdoctoral Researcher and D.B. a Research Director at F.R.S.-FNRS.

L.W. acknowledges the National Natural Science Foundation of China (Grant Nos. 21922305 and 21873080).

I.W., K.C., and J.M.H. acknowledge support from the Marsden Fund New Zealand, through a fast-start grant to K.C. and a standard grant to J.M.H. and K.C..

We thank Diamond Light Source for access and support in use of the electron Physical Science Imaging Centre (Instrument E02 and proposal number MG25140).

**Author contributions:** T.F., I.M., A.R. and R.H.F. conceived of the project. T.F. carried out most of the experimental work to synthesize and structurally characterize the NFs, and T.F., Y.Z., L.R.M., G.R.W., and I.M. discussed the results. A.J.S. and D.P. prepared and characterized the films used for spectroscopy. S.M.C and T.J.A.S performed the SED measurements and analyzed the data. A.J.S performed the TAM measurements under the guidance of J.S.. A.J.S. and D.P. performed the fs-TA measurements. Z.A.G. performed the PDS measurements. I.W. performed the fs-PL measurements under the supervision of K.C.. L.W. and D.B. developed the simulation model. S.P. performed the theoretical calculations and simulations under the supervision of D.B.. T.F. and J.D.G.H. performed the confocal scanning laser microscopy (CLSM). A.J.S. analysed the spectroscopy and microscopy data. A.J.S., D.P. and A.R. interpreted the spectroscopy and microscopy data. A.J.S., A.R., T.F., I.M., and D.B. co-wrote the manuscript with input from all authors.

**Competing interests:** Authors declare no competing interests.

**Data and materials availability:** The data underlying all figures are publicly available from the University of Cambridge repository at DOI: 10.17863/CAM.71186

# Figures

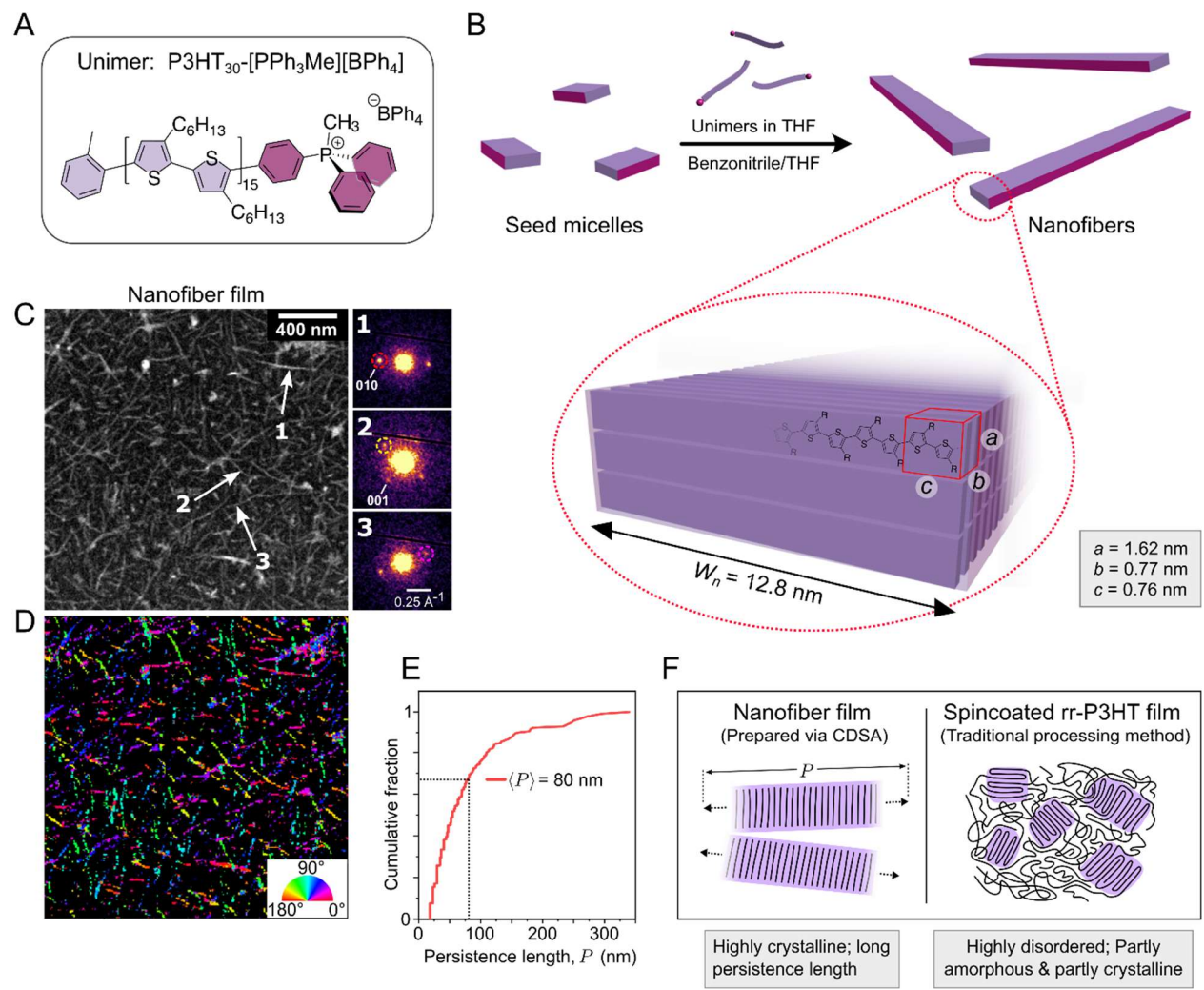

**Fig. 1. System of interest: rr-P3HT-based fiber-like micelles (i.e. nanofibers).** (**A**) unimer component of nanofibers (**B**) schematic of living crystallization-driven self-assembly (CDSA) of nanofibers, with an illustration of the unit cell calculated from powder X-ray diffraction. (**C**) virtual annular dark field STEM micrograph of film of nanofibers with diffraction patterns at the marked points. (**D**) corresponding orientation map. Colour scale represents the orientation of the detected Bragg spots associated with the π-π stacking along the nanofiber. (**E**) corresponding cumulative distribution (per area) of the lengths of the continuous domains that exhibit the same orientation i.e., the "persistence length". The mean persistence length of 80 nm indicates the chains in the nanofibers are highly aligned over large distances in the films. (**F**) Schematic of microstructure of nanofiber films versus spincoated rr-P3HT (a traditional processing method), highlighting how living CDSA results in a superior microstructure with alignment of polymer chains over large distances, while spincoating rr-P3HT results in a mixture of amorphous and crystalline domains, with limited long-range order.

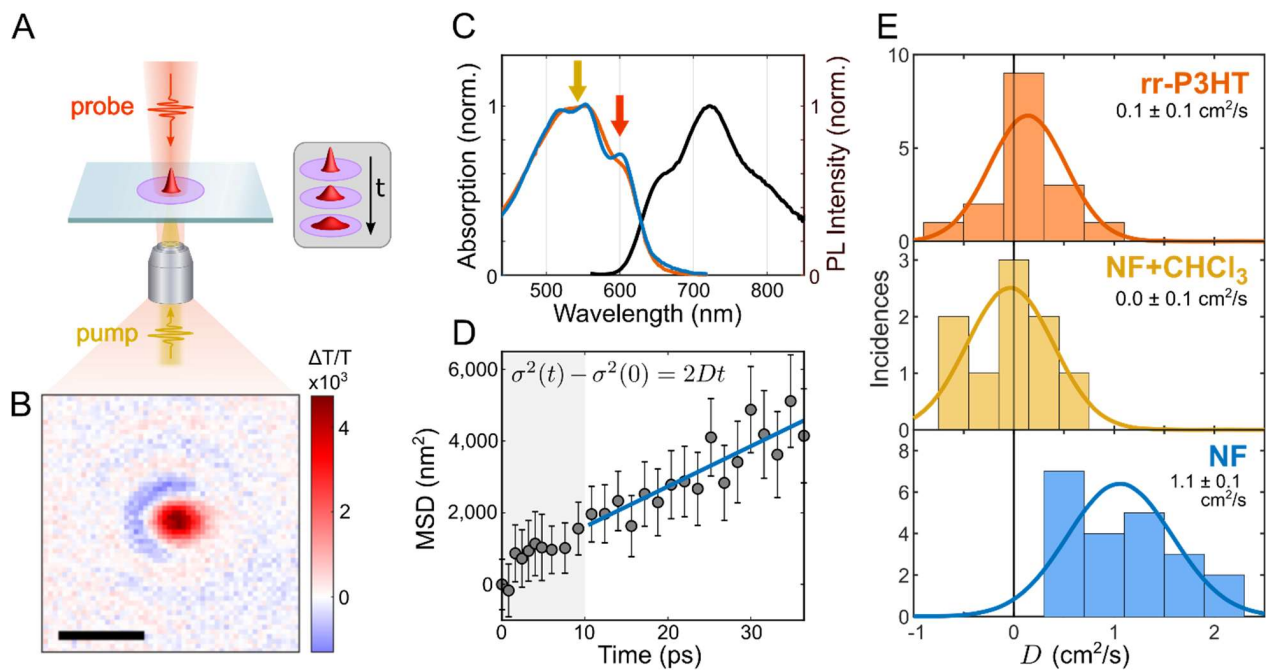

**Fig. 2. Transient-absorption microscopy of nanofiber film and controls.** (**A**) schematic of TAM experiment, with emphasis on expansion of Gaussian profile due to exciton diffusion. (**B**) Example of $\Delta T/T$ image close to time zero measured at 600 nm. (**C**) Normalized absorption (blue) and steady-state photoluminescence spectra (black) of nanofiber film. The absorption of a spincoated rr-P3HT film (orange) matches the nanofibers' absorption due to the identical π-π chain packing motif. The yellow and red arrows indicate the central wavelengths of the pump (540 nm) and probe (600 nm), respectively. (**D**) A typical MSD profile at an average fluence of 4 µJ/cm$^2$ with a linear fit to the region from 10-30 ps to extract $D$. At this later region there is no observed dependence of $D$ with fluence. (**E**) histograms of $D$ from 10-30 ps for multiple measurements on a control spin-coated rr-P3HT film (rr-P3HT), control nanofibers treated with chloroform to destroy their long-range order (NF+CHCl$_3$), and the pristine nanofiber (NF) film. Inset are mean value of $D$ in each case. The nanofibers exhibit a large diffusion constant of $D_{NF} = 1.1 \pm 0.1$ cm$^2$/s, while the controls both show behaviour consistent with a small diffusion constant ($< 0.1$ cm$^2$/s). The clear difference between the nanofibers and controls highlights how the nanofibers' large exciton diffusivity is a function of their superior morphology/energetic order. Note that the spread in the values of $D$ are primarily due to random error. Uncertainties are the standard deviation over $\sqrt{N}$, with $N$ being the number of measurements.

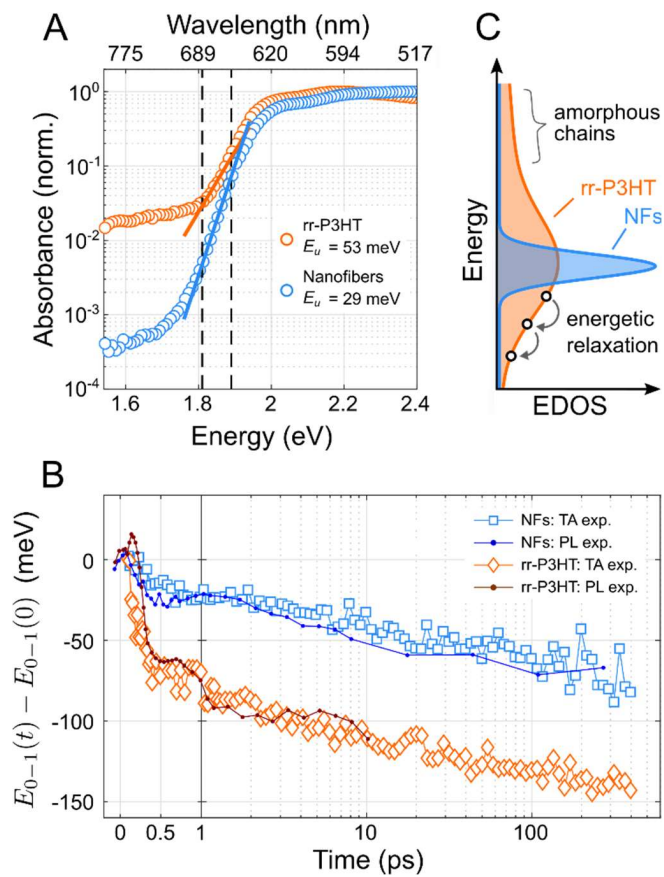

**Fig. 3. Energetic order in nanofiber films versus spincoated rr-P3HT.** (**A**) Photothermal deflection spectroscopy (PDS) reveals that the NFs have an Urbach energy, $E_u$, of $29 \pm 1$ meV, almost half that of spin-coated rr-P3HT films, indicating the NFs are much more energetically ordered. The NFs also have on average over 40× less absorption of low-energy (<1.7 eV) trap states. $E_u$ is found by linear fitting to the regions enclosed by the dashed lines. (**B**) Red-shifting of 0-1 vibronic emission peak (see S.I. for fitting details) independently measured by two techniques: fs-TA and TRPL. Red-shifting indicates relaxation in a disordered EDOS, and accordingly, there is substantially less relaxation in the nanofibers in comparison to rr-P3HT. (**C**) Schematic of EDOS after photo-excitation, illustrating how that in direct contrast to rr-P3HT, the NFs possess a tight EDOS due to energetic order which inhibits the energetic relaxation of excitons. This restricted relaxation promotes the spatial diffusion of excitons as neighbouring sites are more likely to be thermally accessible.

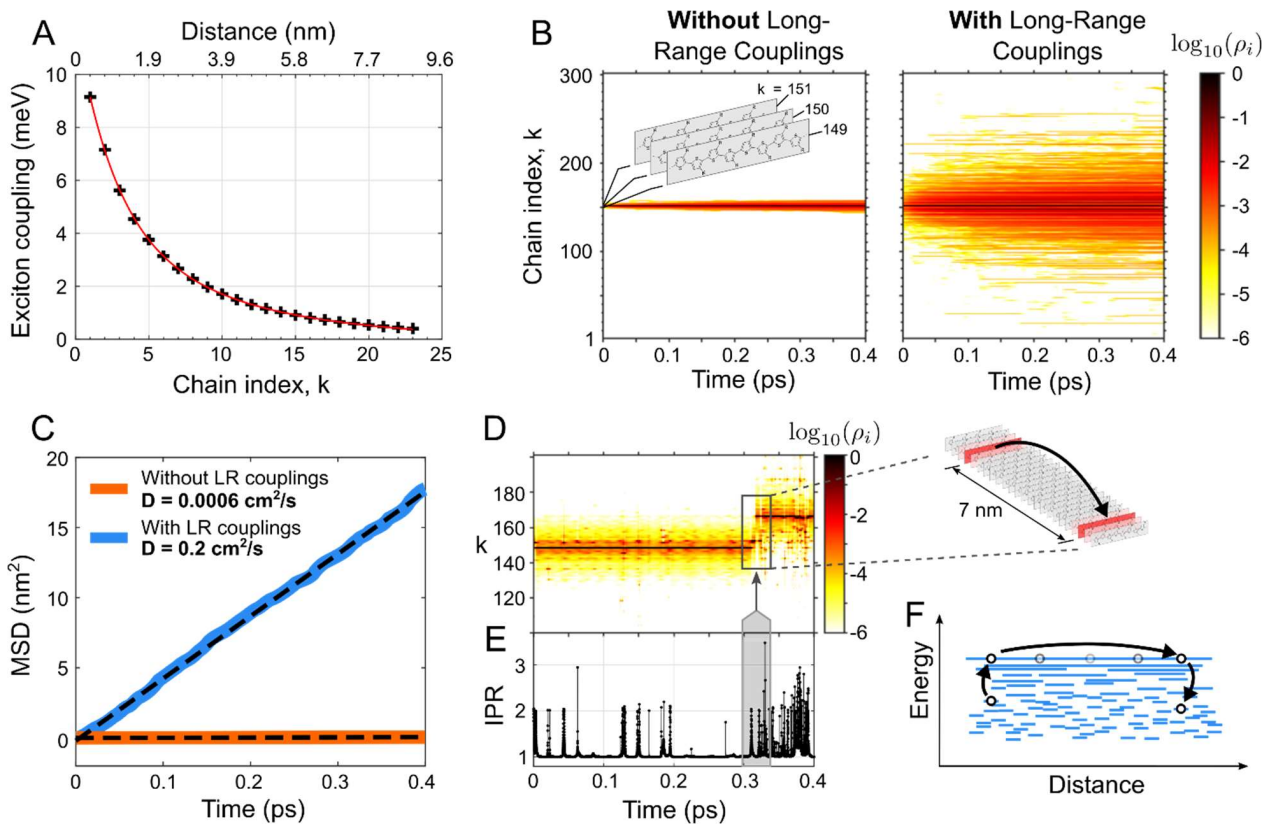

**Fig. 4. Simulations of exciton diffusion.** (**A**) INDO/SCI calculations of the Coulombic exciton couplings in a 24-chain stack, with a bi-exponential fit. (**B**) Sum of many individual surface hopping simulations of exciton diffusion in a 1D stack (i.e. trajectories) with and without long-range exciton couplings. The colour scale indicates the log of the normalized total exciton density. (**C**) The associated MSD derived from (**B**). The MSD evolves linearly, indicating an equilibrium regime is reached. Linear fits (dashed) give diffusion constants of $D = 0.22$ cm$^2$/s and $6 \times 10^{-4}$ cm$^2$/s with and without the inclusion of long-range exciton couplings respectively, which demonstrates how long-range couplings dramatically increase diffusion. (**D**) Exciton density over a single representative trajectory, with (**E**) the associated IPR. The IPR remains mostly low (~1) indicating that the exciton is mostly localized on a single chain, but occasionally (as indicated in the highlighted region) the IPR increases as the exciton transiently occupies higher-energy delocalized states, and when it does the exciton tends to move large distances by 'surfing' along the EDOS as shown by the inset. (**F**) A schematic illustrating how excitons can transiently occupy high-energy delocalized states to move large distances. These delocalized states are responsible for the majority of the diffusion, a fact which is emphasized by making the comparison to the case of no long-range couplings, where no such delocalized states are occupied, and the diffusion is severely restricted.

**Supplementary Materials**

Supplementary Text
Figures S1-S31
Table S1

# Supplementary Materials for

## Efficient Energy Transport in an Organic Semiconductor Mediated by Transient Exciton Delocalization

Alexander J. Sneyd, Tomoya Fukui, David Paleček, Suryoday Prodhan, Isabella Wagner, Yifan Zhang, Jooyoung Sung, Sean M. Collins, Thomas J. A. Slater, Zahra Andaji-Garmaroudi, Liam R. MacFarlane, J. Diego Garcia-Hernandez, Linjun Wang, George R. Whittell, Justin M. Hodgkiss, Kai Chen, David Beljonne\*, Ian Manners\*, Richard H. Friend, and Akshay Rao\*

\* Correspondence to: ar525@cam.ac.uk, imanners@uvic.ca, david.beljonne@umons.ac.be.

**This PDF file includes:**

Supplementary Text
Figs. S1 to S31
Table S1

# Supplementary Text

## 1 Synthesis of nanofibers
### 1.1 Synthesis of P3HT$_{30}$-[PPh$_3$Me][BPh$_4$]

**P3HT$_{30}$-[PPh$_3$Me]I** is synthesized according to the literature(*21*). To remove the redox-active iodide counteranion, the THF solution of **P3HT$_{30}$-[PPh$_3$Me]I** (200 mg), THF/MeOH mixed solution of potassium tetraphenyl borate (1.0 g, 2.8 mmol, excess) was added slowly under stirring. After 1 h, formed precipitates were collected by centrifuge and the obtained solid was washed by methanol. Then, the solid was dissolved in chloroform and washed by water. The solution was condensed under vacuum and then poured into hexane. The formed precipitates were collected by centrifuge and dried at 40 ºC under vacuum for 24 h. **P3HT$_{30}$-[PPh$_3$Me][BPh$_4$]** was obtained as purple solid (162 mg, Yield: 78 %).

$^1$H NMR (500 MHz, CDCl$_3$) δ 7.81–7.78 (m, aromatic), 7.47–7.43 (m, aromatic), 7.48–7.43 (m, aromatic), 6.98 (s, thiophene), 6.81–6.79 (m, aromatic), 3.49 (d, P-*CH$_3$*), 2.81(t, *J* = 7 Hz, -*CH$_2$*-), 2.49 (s, -*CH$_3$*), 1.77–1.71 (m, -*CH$_2$*-), 1.47–1.41 (m, -*CH$_2$*-), 1.38–1.32 (m, -*CH$_2$*-), 0.92 (t, -*CH$_3$*). $^{31}$P NMR (202 MHz, CDCl$_3$) δ 20.8.

The number-average degree of polymerization (DP$_n$) was determined using both $^1$H NMR and MALDI-TOF-MS. To evaluate DP$_n$ based on NMR analysis, we used a peak at 2.49 ppm (C*H$_3$*- group on tolyl group) as a standard, and calculated the ratio of integrals of each peaks. The obtained ratio was consistent with the value obtained by MALDI-TOF-MS.

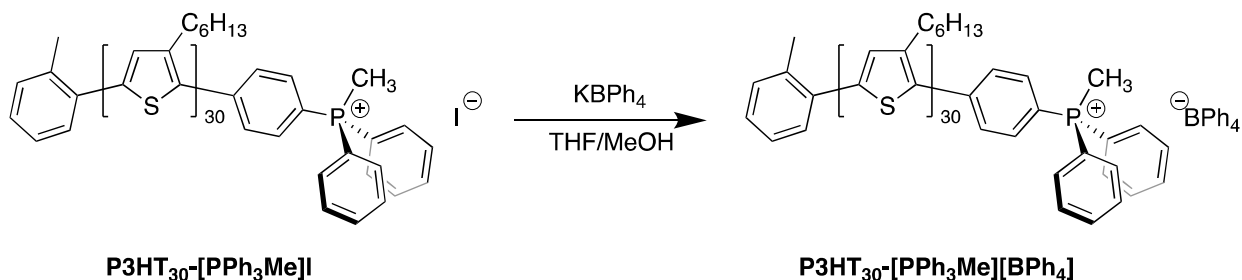

**Fig. S1: Synthesis Schematic.** Schematic of synthesis of P3HT$_{30}$-[PPh$_3$Me][BPh$_4$] from P3HT$_{30}$-[PPh$_3$Me][I]

### 1.2 Experimental procedures for the self-assembly

*Preparation of fiber-like micelles of P3HT$_{30}$-[PPh$_3$Me][BPh$_4$]:*

**P3HT$_{30}$-[PPh$_3$Me][BPh$_4$]** was dissolved in benzonitrile/THF mixture (benzonitrile:THF = 8:2, Conc. 0.5 mg/mL) at 100 °C for 1 h. Subsequently, the hot solution was slowly cooled to room temperature. The solution was aged under stirring at room temperature for 12 h.

*Preparation of seed micelles of P3HT$_{30}$-[PPh$_3$Me][BPh$_4$]:*

The solution of the polydisperse fiber-like micelles (Conc. = 0.5 mg/mL) was degassed with N$_2$ for 15 min. The polydisperse fiber-like micelles were fragmented by applying the sonication under N$_2$ at 10 ºC for 2 h to yield short micelles. The short micelles were thermally annealed at 40 ºC for 30 min and then cooled to room temperature. The length of resulting seeds was evaluated by TEM analysis.

*Living CDSA for P3HT$_{30}$-[PPh$_3$Me][BPh$_4$] at low concentration*

The solution of the seeds (Conc. = 0.5 mg/mL) was diluted by benzonitrile/THF mixture (benzonitrile:THF = 8:2) to the appropriate concentration. The solution of seed micelles was warmed at 35 ºC for 5 min. To the seed micelles, appropriate amounts of unimers in THF were added (final Conc. = 0.2 mg/mL). After the mechanical shaking for 10 sec, the solutions were aged for 12 h. TEM analyses evaluated the length of the resulting micelles. By varying the unimer to seed ratio samples of uniform nanofibers with lengths between ca. 250 – 2100 nm were prepared.

Note that the concentration of the resulting micelles was too high for the TEM analysis. Therefore, the solution was diluted to 0.02 mg/mL using benzonitrile/THF mixture.

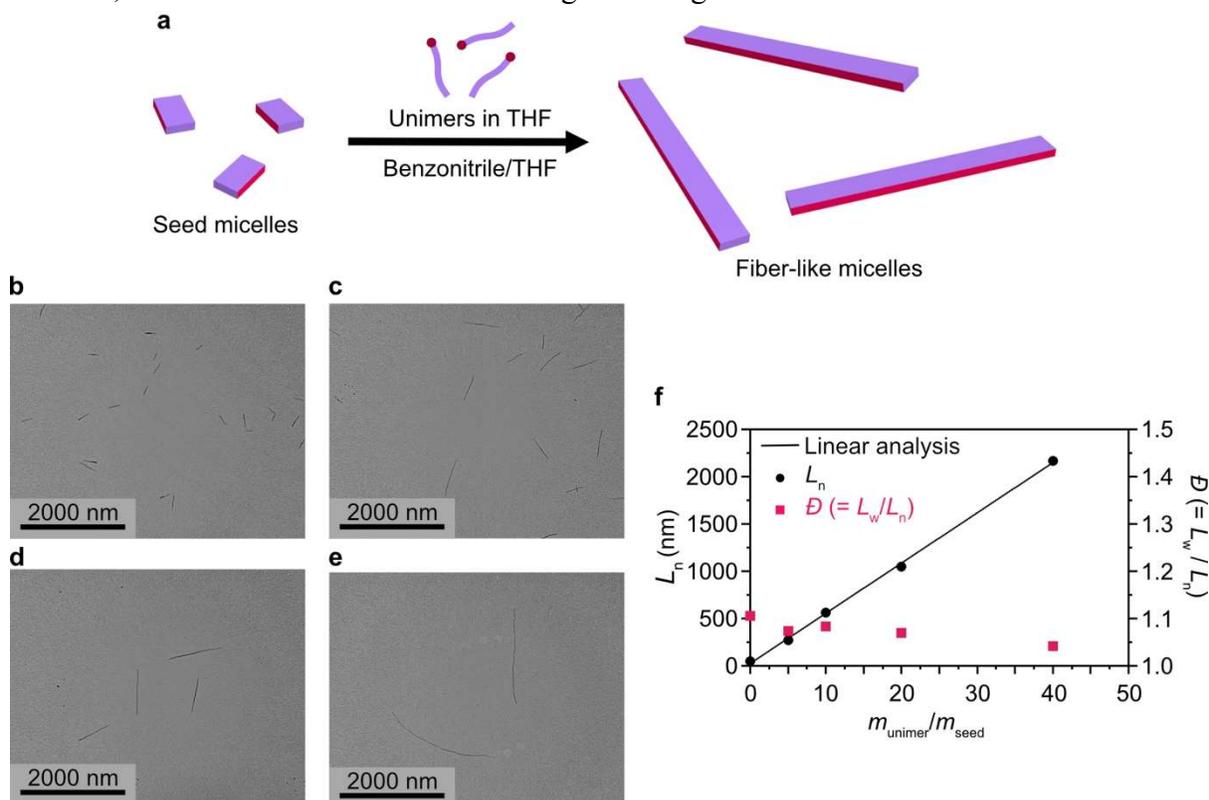

**Fig. S2: Low-concentration living CDSA.** (**a**) Schematic representation of living CDSA at **low concentration**. (**b-e**) TEM images of micelles of **P3HT$_{30}$-[PPh$_3$Me][BPh$_4$]** formed by living CDSA; unimer to seed ratio $m_{unimer}/m_{seed}$ = 5 (**b**), 10 (**c**), 20 (**d**), and 40 (**e**). (**f**) Number-average length ($L_n$) of micelles and polydispersity index ($Đ$) of **P3HT$_{30}$-[PPh$_3$Me][BPh$_4$]** vs unimer to seed ratio ($m_{unimer}/m_{seed}$).

*Living CDSA for P3HT$_{30}$-[PPh$_3$Me][BPh$_4$] at high concentration:*

The solution of the seed micelles (conc. = 0.5 mg/mL) in benzonitrile/THF mixture (benzonitrile:THF = 8:2) was warmed at 35 ºC for 5 min. To the seed micelles, appropriate amounts of unimers in THF were added. After the mechanical shaking for 10 sec, the solutions were aged for 12 h. THF was evaporated by gentle N$_2$ gas flow for several hours (final conc. = 1.0 mg/mL). By varying the unimer to seed ratio samples of uniform nanofibers with lengths between ca. 250 – 900 nm were prepared. TEM analyses evaluated the length of the resulting micelles. Note that the concentration of the resulting micelles was too high for the TEM analysis. The solution was therefore diluted to 0.02 mg/mL by using benzonitrile/THF mixture.

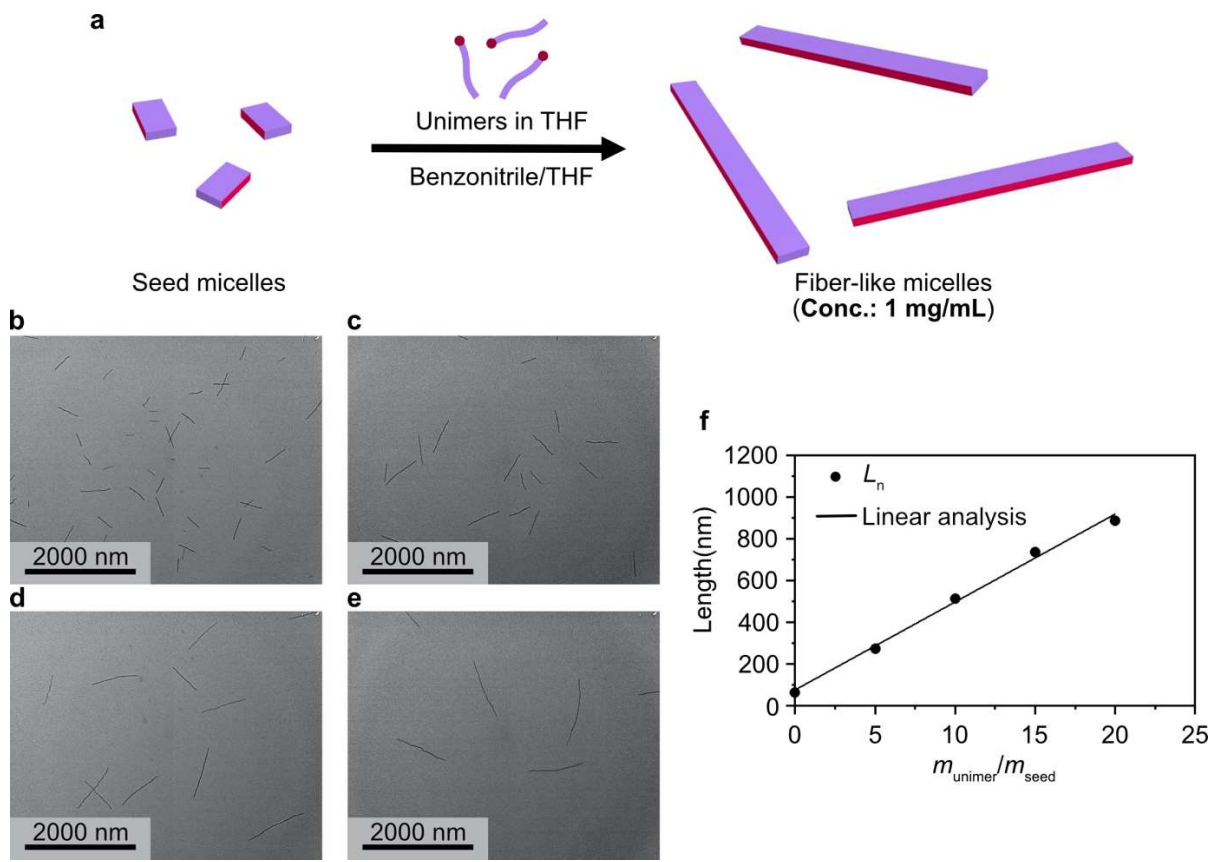

**Fig. S3**: **High-concentration living CDSA** (**a**) Schematic representation of living CDSA at **high concentration**. (**b-e**) TEM images of micelles of **P3HT$_{30}$-[PPh$_3$Me][BPh$_4$]** formed by living CDSA; unimer to seed ratio $m_{unimer}/m_{seed}$ = 5 (**b**), 10 (**c**), 20 (**d**), and 40 (**e**). (**f**) Number-average length ($L_n$) of micelles of **P3HT$_{30}$-[PPh$_3$Me][BPh$_4$]** vs unimer to seed ratio ($m_{unimer}/m_{seed}$).

## 2 Structural characterization of nanofibers

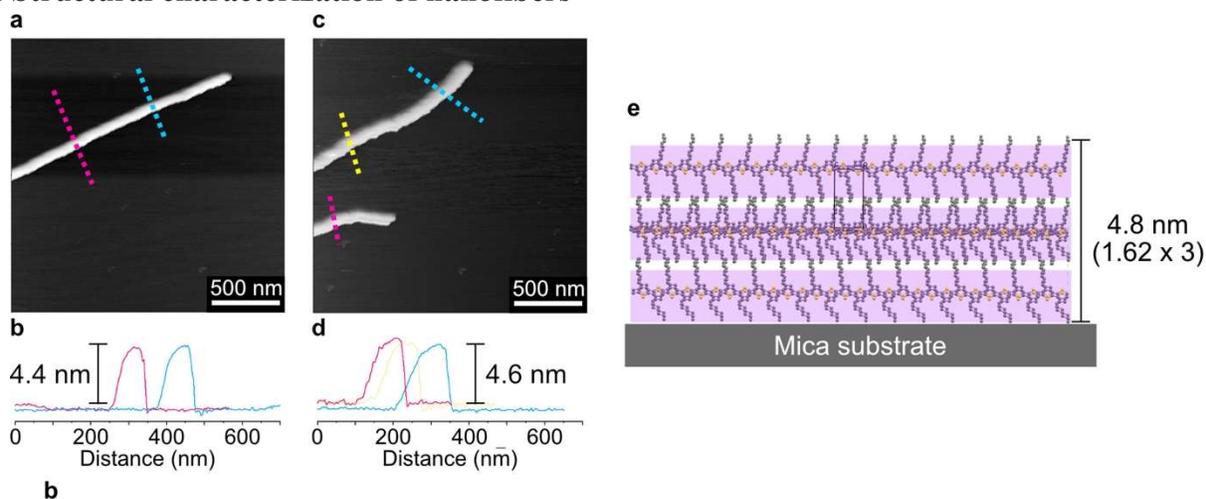

**Fig. S4: AFM of nanofibers.** (**a-d**) AFM images (**a** and **c**) and height profile (**b** and **d**) of nanofibers derived from **P3HT$_{30}$-[PPh$_3$Me][BPh$_4$]**. (**e**) Schematic representation of height profile of fiber-like micelles derived from **P3HT$_{30}$-[PPh$_3$Me][BPh$_4$]**. Note that the widths obtained by AFM are not accurate due to tip deconvolution effects. Nanofiber widths were determined by TEM (12.8 nm) and heights (4.5 nm) by AFM.

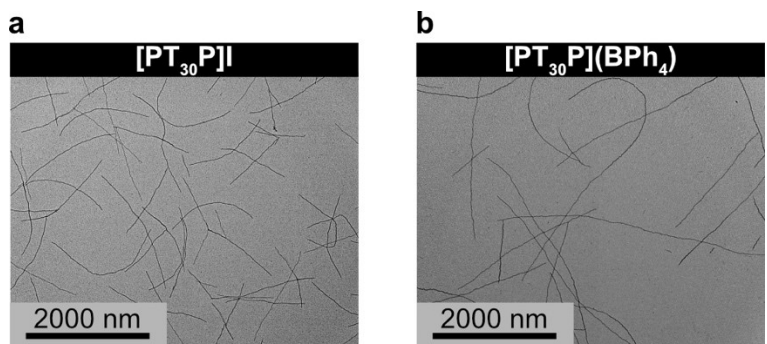

**Fig. S5: TEM of nanofibers,** TEM images of nanofibers consist of **P3HT$_{30}$-[PPh$_3$Me]I** (**a**) and **P3HT$_{30}$-[PPh$_3$Me][BPh$_4$]** (**b**) on the carbon-coated Cu grid.

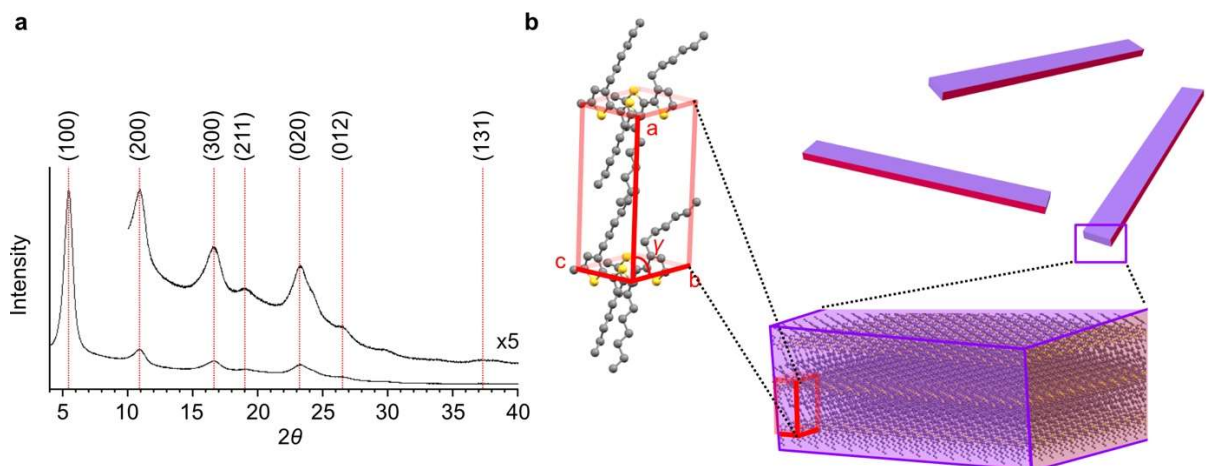

**Fig. S6: Power XRD of nanofiber film with illustration of unit cell.** (**a**) Powder XRD for the films of the nanofibers derived from **P3HT$_{30}$-[PPh$_3$Me][BPh$_4$]**. (**b**) Schematic representation of the unit cell of the nanofibers. The diffraction pattern was almost identical to that in the literature (*49*).

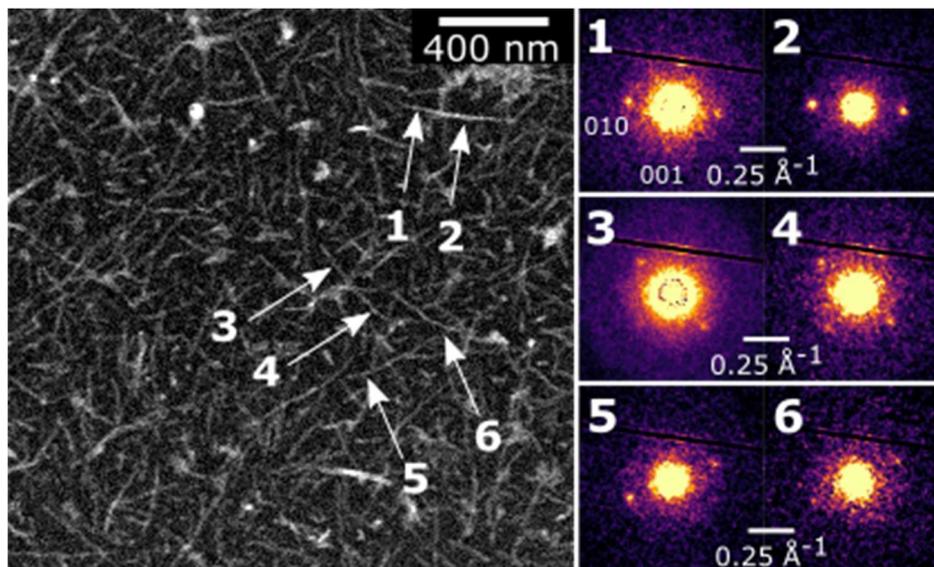

**Fig. S7: Scanning electron diffraction (SED) of nanofiber film.** (Left) Virtual annular dark field STEM micrograph showing nanofiber film morphology on a TEM grid and (right) diffraction patterns collected at areas marked by the white arrows. The dark line in the diffraction patterns arises from a spacing in the Merlin-Medipix chip readout array.

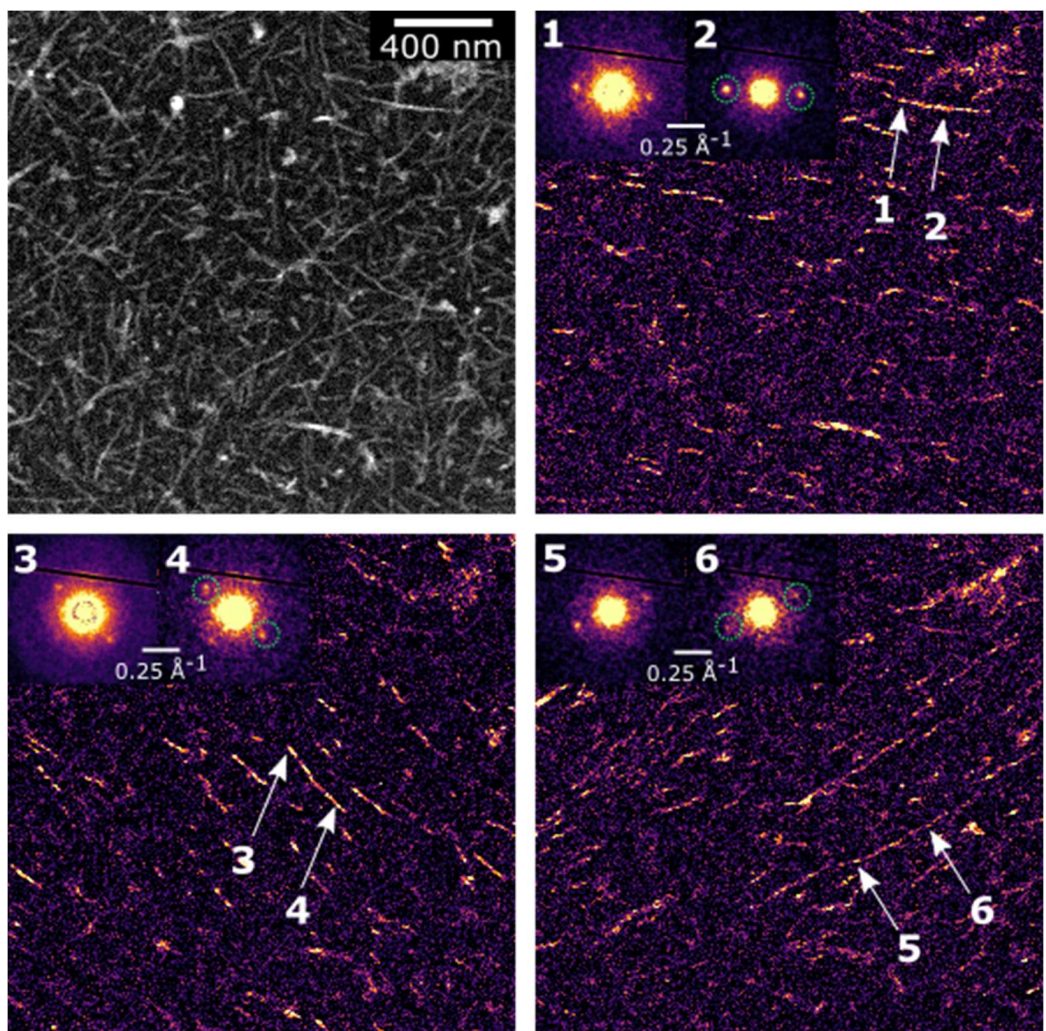

**Fig. S8: SED of nanofiber film examining orientation of diffraction spots.** (a) Virtual annular dark field STEM micrograph showing nanofiber film morphology on a TEM grid. (b)-(d) Virtual dark field STEM images mapping the crystalline distribution and orientation along the nanofibers, formed from selected scattering vectors corresponding to Bragg diffraction spots obtained in two-dimensional diffraction patterns recorded at each pixel in (a). The insets show diffraction patterns collected at areas marked by the white arrows with the virtual aperture positions indicated by green dashed lines. The dark line in the diffraction patterns arises from a spacing in the Merlin-Medipix chip readout array.

2.1 Extraction of persistence length from scanning electron diffraction

    The distribution of aligned domain sizes, that is, the persistence length of the regions detected with strong Bragg diffraction assigned to the π-π stacking along the nanofiber, was analysed by segmentation of a stack of VDF-STEM images calculated at 1° increments using a pair of virtual apertures at the Bragg spots corresponding to the π-π stacking distance. This definition of aligned domain size or persistence length of the crystalline material along the nanofiber assumes the π-π stacking is near-perpendicular to the electron beam trajectory (in/out of the plane of the image) and is accounted for by in-plane rotation of the π-π stacking direction

only. The definition does not account for (undetectable) crystalline domains where the nanofibers rotate out-of-plane, as this rotation causes deviations from the Bragg condition.

To calculate the aligned domain lengths, the VDF-STEM image stack was thresholded and segmented in Scikit-image, an open-source Python package for image processing. Thresholds were selected by evaluating a series of thresholds and selecting a balance of noise removal, while avoiding excessive reduction in size of the domains apparent in the VDF images. Each domain was labelled based on pixel-wise connectivity in the three-dimensional (x,y,θ) stack of VDF-STEM images. This segmentation was then projected along the rotation dimension for each separated domain. Domains corresponding to single pixels were removed, and the segmentation was inspected to remove any aggregates with extreme areas. The separated domains were analysed using ImageJ particle analysis. Similar measurements were also obtained using Scikit-image functions. The area and length of the domains (defined as the diagonal of the bounding box) were recorded for each particle in (x,y) space of the image. This metric for length is justified for fiber-like geometries though may over-estimate length for rounded features or under-estimate length for fibers with significant curvature within the bounding box. Similar results were obtained by examining equivalent ellipse major axes in Scikit-image.

A cumulative distribution was then constructed by sorting the separated particles by detected persistence lengths, with each length weighted by the detected area of the associated domain. Weighting the distribution by area (as opposed to by-number) helps to account for nanofibers which may be 'stuck' closely together and are not easily identifiable as separate nanofibers. The mean of this 'per area' distribution is given as:

$$\bar{L}_A = \frac{\sum_j p_j L_j}{\sum_j p_j} = \frac{\sum_j p_j L_j}{A} \qquad (2.1)$$

where $\bar{L}_A$ is the mean length in nanometers, $p_j$ is the number of pixels (units of area) of unique domain $j$ with length $L_j$. This expression simplifies to the area-weighted length over the total area $A$.

The $\bar{L}_A$ value can be interpreted as the average length for any given unit of area selected from the aligned nanofiber domains, i.e. any selected crystalline pixel would on average be part of a aligned domain that is $\bar{L}_A = 80$ nm in length. In the main text we take $\bar{L}_A$ to be the mean persistence length i.e.

$$\langle P \rangle = \bar{L}_A \qquad (2.2)$$

To illustrate the physical meaning of the persistence length in relation to overall length of the nanofibers, please refer to fig. S9:

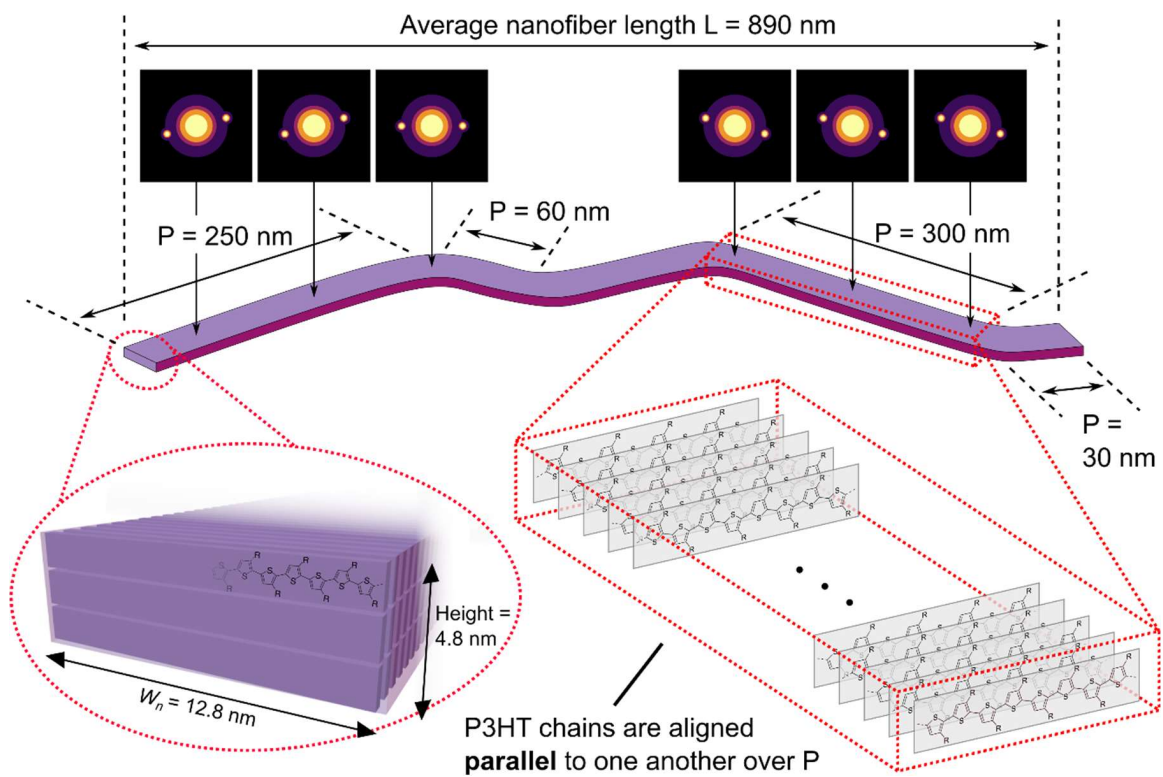

**Fig. S9: Schematic of nanofiber, persistence length, and information gained from SED.**
Illustration of the single P3HT nanofiber in the film (not to scale), with annotations noting various hypothetical persistence lengths P in the nanofiber. Bends are exaggerated to make the definition of P clearer. Schematic examples of corresponding diffraction patterns that would be recorded in scanning electron diffraction (SED) are annotated at various points along the nanofiber. The oval inset gives the width and height of the nanofibers. Note that the average length of the nanofibers found from TEM analysis is 890 nm (with a polydispersity index of 1.08). The rectangular inset illustrates how in an aligned domain (which is in case has a large hypothetical persistence length of 300 nm), the P3HT chains are aligned parallel to one another over this entire length, enabling strong long-range dipole couplings between different chains. Note that the average persistence length calculated from our data is 80 nm, with lengths up to 300 nm recorded. Since the chains are aligned parallel to one another this causes the Bragg diffraction spots to be recorded at the same angle, which allows for identification of these aligned domains in the first place (as discussed in Figure 1 of the main text and section 2 of the SI).

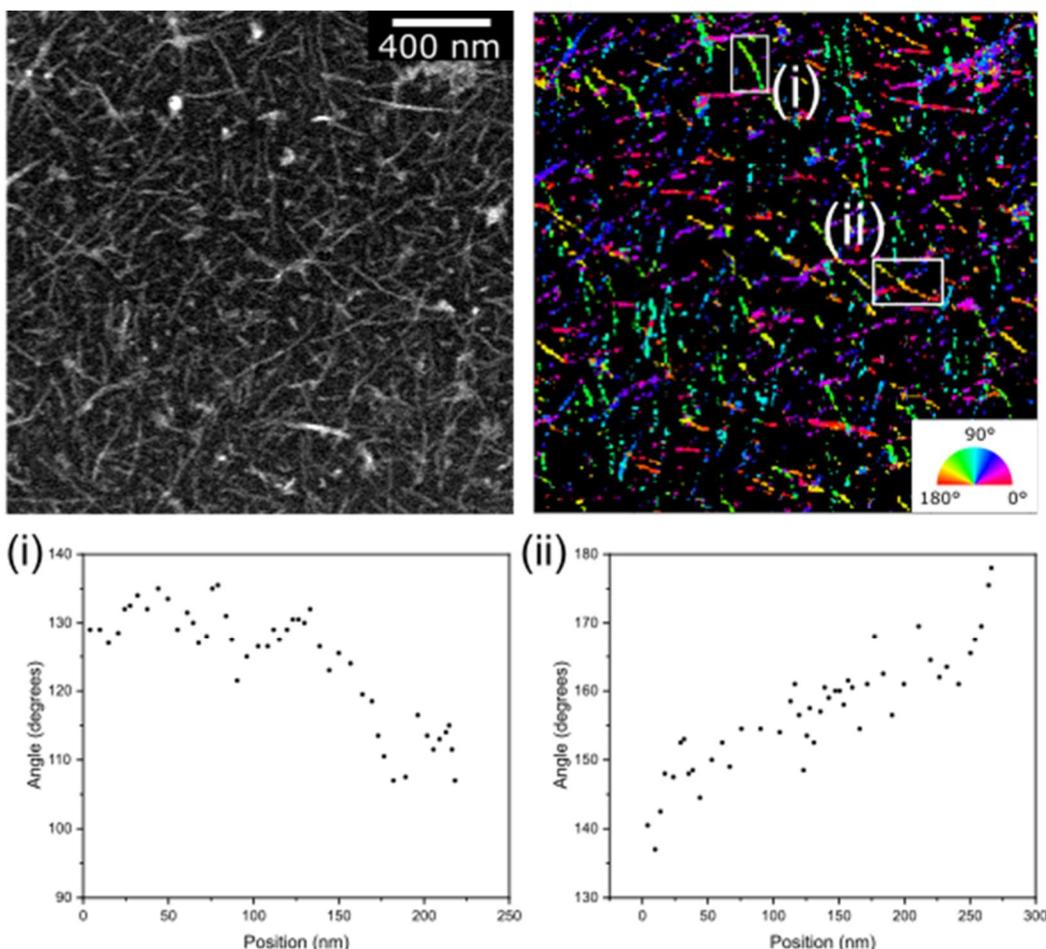

**Fig. S10: Evaluation of the in-plane rotation of the π-π stacking direction along bent nanofibers**. (Left) ADF-STEM micrograph and (right) the corresponding orientation map, masked according to the areas segmented for length analysis. The color scale represents the orientation of the detected Bragg spots associated with π-π stacking along the nanofiber. The inset shows the upper half-plane color-wheel and the corresponding orientation angles. Two fibers were selected, corresponding to the orientation plots labelled (i) and (ii), respectively.

2.2 In-plane rotation of bent nanofibers

The in-plane rotation of visibly bent nanofibers was assessed by tracking the Bragg diffraction spots associated with the π-π stacking along the nanofiber (fig. S10). Fibers showing significant bending were selected to capture limiting case behaviour. The orientation of the π-π stacking direction was determined as the maximum in the VDF image stack for each pixel. The segmentation used for domain length was applied as a mask to distinguish the highly crystalline pixels only. For selected fibers, pixels not associated with the fiber axis were excluded, and the angles detected along the fiber were plotted as a function of position. The position axis is defined as the distance from the tip of the nanofiber (top left). Changes of approximately 10° within 50 nm distance were detected for these fibers, aligned with the bending behaviour apparent in the fiber morphology.

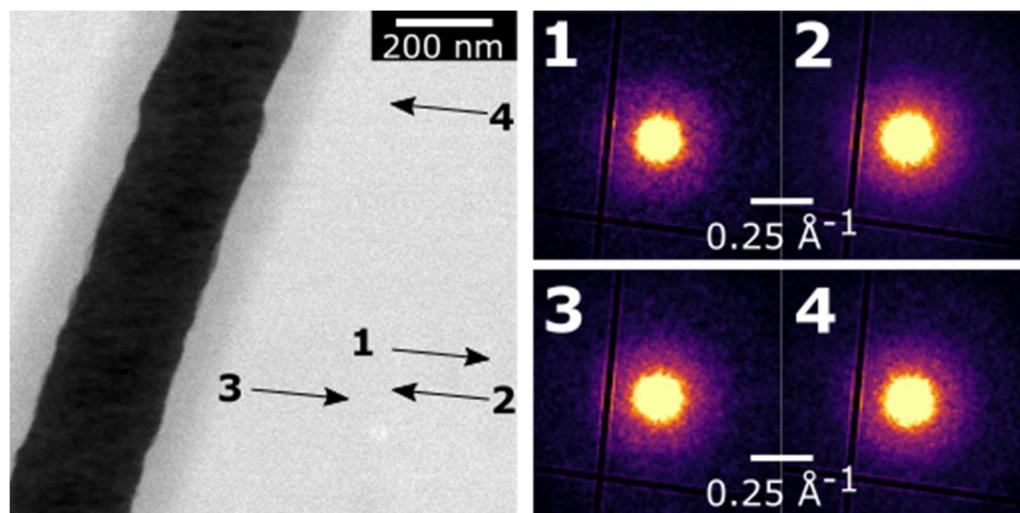

**Fig. S11: SED of pieces of rr-P3HT film floated onto TEM grid**. Dark region is the edge of a film section. Diffraction patterns showed no detectable Bragg spots from areas examined based on faint contrast fluctuations in a VDF image stack selecting for scattering vectors at the π-π stacking as for the nanofiber samples. SED data were acquired under an identical electron optical configuration and electron fluence as for nanofiber samples.

## 3 Ensemble steady-state spectroscopy

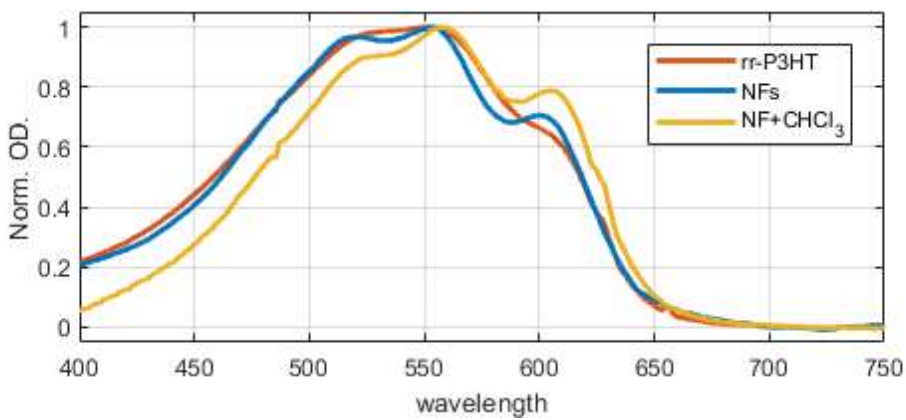

**Fig. S12: UV-Vis absorption of a spin-coated rr-P3HT thin film, nanofiber film (NFs), and the nanofiber film treated with CHCl$_3$ to disrupt the morphology (NF+CHCl$_3$).** The absorption of the rr-P3HT and NFs match each other very well, with the main difference being that the NFs have sharper vibronic peaks, which is an indicator that the nanofibers have less inhomogeneous broadening, and are thus less disordered. When treated with CHCl$_3$, the absorption of the NFs does not change much, demonstrating that this sample is a valid control system for transient-absorption measurements.

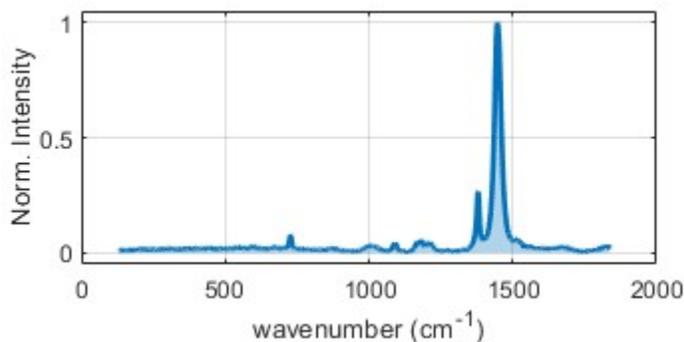

**Fig. S13: Raman spectrum of nanofibers.** Normalized Raman spectrum of NF film with 532 nm excitation, demonstrating the strong coupling of the singlet electronic state to the 1450 cm$^{-1}$ mode corresponding to the aromatic C=C stretch. The PL background has been subtracted to reveal the vibrational modes more clearly.

## 4 Confocal laser scanning microscopy

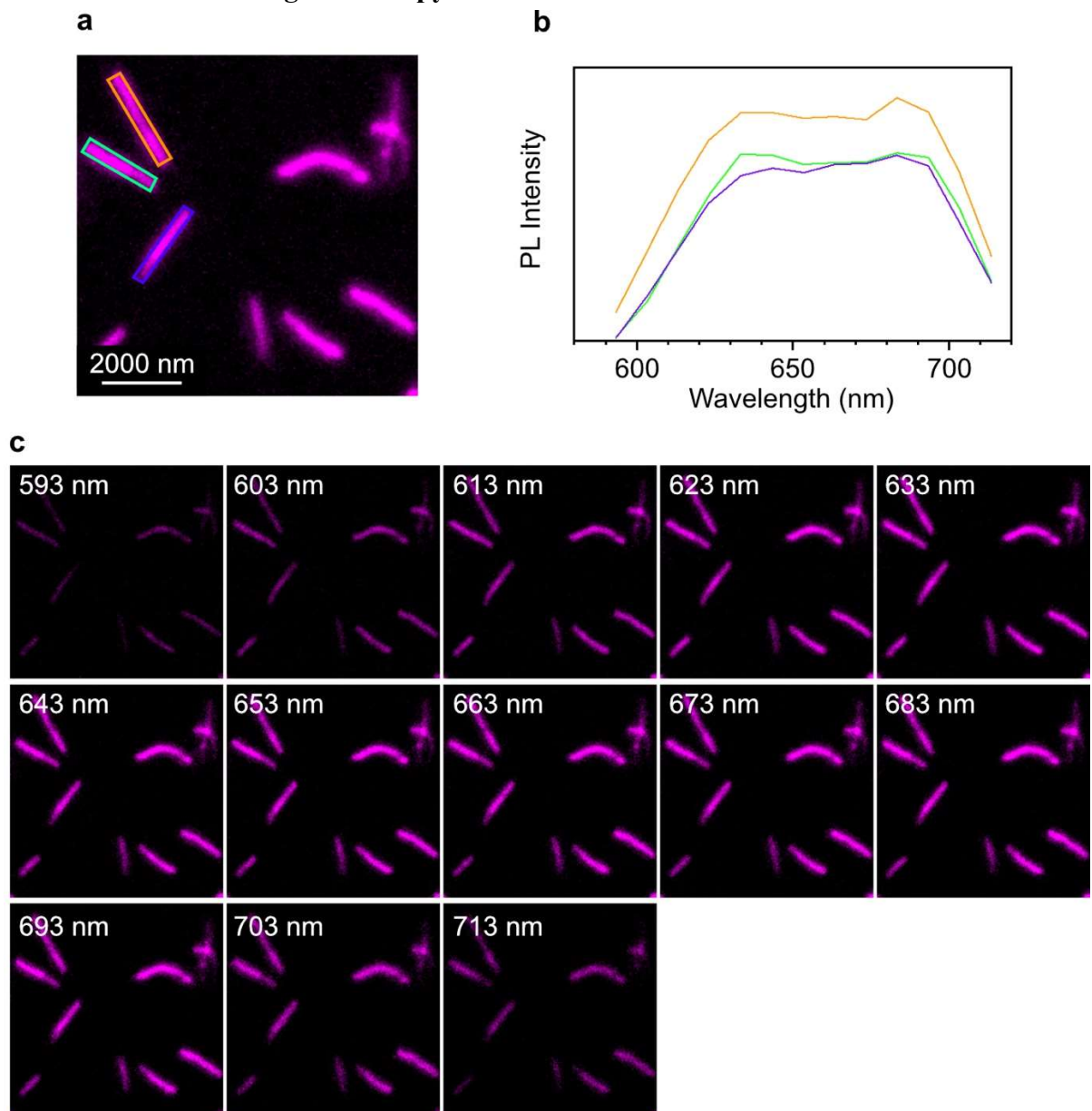

**Fig. S14: CLSM of nanofibers.** (**a**) CLSM image of the nanofibers of **P3HT$_{30}$-[PPh$_3$Me][BPh$_4$]** in benzonitrile/THF mixture (Conc. = 0.02 mg/mL). (**b**) Photoluminescence spectrum of a single nanofiber of **P3HT$_{30}$-[PPh$_3$Me][BPh$_4$]** in the area indicated in (**a**). (**c**) CLSM images of micelles of **P3HT$_{30}$-[PPh$_3$Me][BPh$_4$]** detected at each wavelength.

# 5 Ultrafast transient-absorption spectroscopy (fs-TA)

## 5.1 Assignment of species in fs-TA

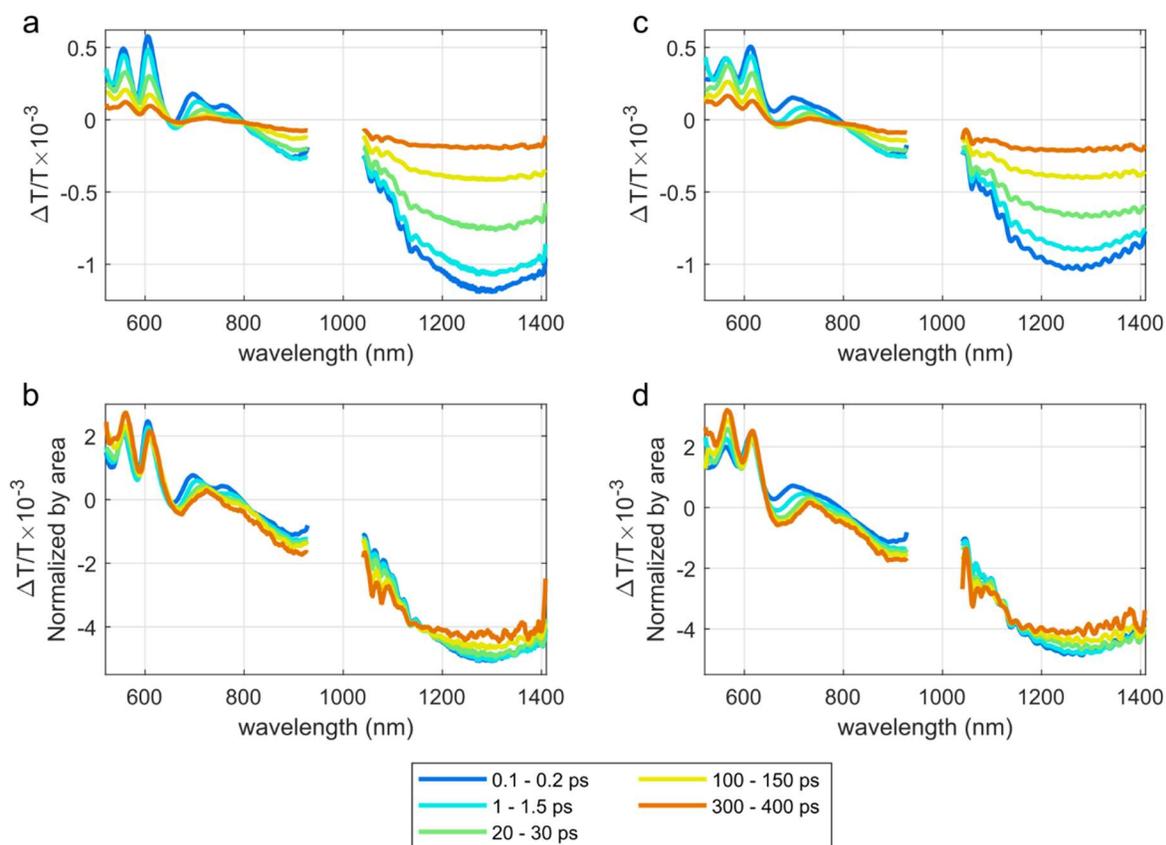

**Fig. S15: Basic fs-TA of nanofiber film and rr-P3HT control film.** Representative fs-TA spectra under ~15 fs 540 nm broadband excitation with parallel pump/probe polarization, with normalized spectra also given to aid visual interpretation. (**a,b**) drop-casted film of nanofibers excited with a fluence of 1.0 $\mu J/cm^2$. (**c,d**) spin-coated rr-P3HT control film excited with a fluence of 0.52 $\mu J/cm^2$. A Savitzky-Golay filter is applied to aid visual interpretation.

Figure S15 gives representative fs-TA spectra for a drop-casted nanofiber film (NFs) and spin-coated rr-P3HT (herein referred to as "rr-P3HT"). As demonstrated in fig. S12, the NF's absorption is much same as rr-P3HT's, with all the same features. Therefore, it comes as no surprise both exhibit very similar TA spectra. The features of note include a ground-state-bleach (GSB) signal below 610 nm arising from the pump-induced initial population of excitons. From 700-800 nm we observe a stimulated emission (SE) signal superimposed on a broad photo-induced absorption (PIA) which extends out to the near-IR region with the peak centred at ~1250 nm.

For the NFs, the TA data shown in fig. S15 at low fluence only gives evidence of a single species, as there is no significant change in the TA spectrum over time. We assign this dominant species to singlet excitons, since the TA spectrum exhibits a strong PIA at 1200-1400 nm and a clear Stokes-shifted SE, which are both known to correspond to singlet excitons (*50*). We note that there is a red-shift in the SE over time, and this process is assigned to the gradual tendency for excitons to diffuse to sites at slightly lower energies (*30*). The fact we observe no evidence of polarons (at low fluences) aligns well with what has been previously shown for polaron (i.e. charge) generation in neat rr-P3HT. In particular, their generation from photoexcitation is known to be related to the energy of the excitation (*27*), and photoemission spectroscopy places the energy of free polarons ~0.7 eV above the bandgap (*51*). Our photo-excitation centered at 540 nm is below the threshold for this to happen. Furthermore, Paquin *et al.* have shown that charge generation in neat rr-P3HT proceeds extrinsically due to a driving force, (as opposed to an intrinsic mechanism directly at photoexcitation) (*52*). This driving force is provided by the presence of ordered and disordered domains (*52, 53*). Since the NFs display exceptional crystalline order with no detectable amorphous domains (a situation which is strongly favoured by the slow epitaxial-like nature of the synthetic process), it follows that spontaneous charge formation is highly unlikely.

Instead, it is perhaps more important to consider the generation of polarons from exciton-exciton annihilation (EEA). This is because when one exciton unloads its energy onto another exciton during EEA, that second remaining 'hot' exciton momentarily has twice the energy of an exciton at thermal equilibrium. This excess energy can then allow for the generation of two free polarons in competition with vibrational relaxation back down to the bandgap (*50*). In line with previous observations on rr-P3HT we do indeed observe the slight formation of a state resembling a polaron (i.e. a state which has a PIA peak in the region from 1000-1100 nm) (*27, 50*) as shown in fig. S16. We can estimate the population of polarons by considering the magnitude of the $\Delta T/T$ at 1050-1100 nm and 1300-1400 nm and exploit the fact that excitons and polarons will have different molar absorption coefficients at each region. By using the ratio of the molar absorption coefficients extracted by Guo *et al.* (*50*), we find that the density of polarons remains very small throughout, and never rises above ~4% of the initial density of excitons at highest fluences used in fs-TA (8 $\mu$J/cm$^2$) – see fig S16. We also calculate a polaron yield of 7%, with this yield defined as the proportion of EEA events that produce polarons. This is roughly consistent with previous reports where it was reported that vibrational relaxation dominates over charge generation in rr-P3HT, with the polaron yield estimated at 5%.(*50*) It is also consistent with microwave conductivity experiments which show that even with excitation at 5.2 eV, the quantum yield of polarons is only 7% (*51*). As a result, we must agree with the analysis used by others (*7, 26, 27*), where polaron generation is not considered in the construction of EEA models to extract the diffusion constant, and hot excitons produced from EEA are always assumed to relax vibrationally.

As a final note, due to the much longer lifetime of polarons in comparison to excitons, at long times (~300 ps) at 8 $\mu$J/cm$^2$, the density of polarons begins to match that of excitons, with both densities at about 2.6% of the original exciton density. However, by this time, the densities of both species are so low (~4 × 10$^{16}$ cm$^{-3}$) that secondary interactions between excitons and polarons will be highly unlikely. This probability of secondary-charge-exciton interactions is further reduced by the fact that, statistically-speaking, the carrier density is more spatially-even at long times. We also suggest that polarons will be less mobile than excitons due to the absence of long-range couplings (which we posit allow excitons to move effectively), and so the probability of secondary-charge-exciton interactions is further reduced, and is thus not an important factor in fs-TA for our analysis.

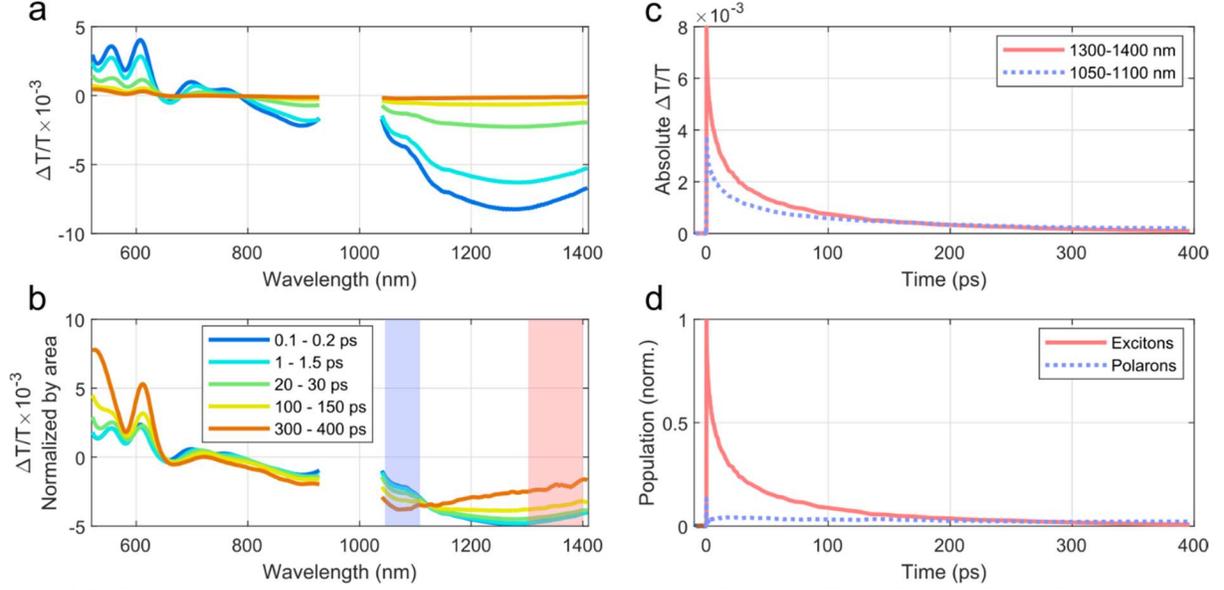

**Fig. S16: Exciton versus polaron populations derived from fs-TA of nanofiber film. (a, b)** Representative fs-TA spectra under ~15 fs 540 nm broadband excitation (parallel pump/probe polarization) of drop-casted film of nanofibers excited with a fluence of 8 $\mu J/cm^2$, with normalized spectra and Savitzky-Golay filter applied to aid visual interpretation. **(c)** The associated $\Delta T/T$ kinetics (anisotropy-corrected) for the regions of 1300-1400 nm and 1050-1100 nm. **(d)** Estimated densities of excitons and polarons from (c), where each trace is normalized by the density of excitons at time zero.

5.2 Red-shift in fs-TA

As previously mentioned, we observe a red-shift in the SE. This is consistent with the red-shift in the PL measured in TRPL. To quantify this red-shift we fit the following to each spectral slice:

$$\frac{\Delta T}{T} = A_{0-1} \text{Exp}\left\{-\frac{(E - E_{0-1})^2}{2\sigma_{0-1}^2}\right\} + A_{0-2} \text{Exp}\left\{-\frac{\left(E - (E_{0-1} - v)\right)^2}{2\sigma_{0-2}^2}\right\} + aE + b \quad (5.1)$$

where $E_{0-1}$ is position of the 0-1 vibronic transition, $v$ is the spacing between $E_{0-1}$ and $E_{0-2}$, $\sigma_{0-i}$ is the standard deviation for the $i$-th vibronic peak, $A_{0-i}$ are the respective amplitudes, and $aE + b$ is a linear background to parametrize the broad overlapping PIA. The fits are performed over the region from 1.46 eV to 1.82 eV. The 0-0 vibronic transition is ignored due to myriad overlapping features. We also note that there is no fluence dependence of the red-shift in the NFs or P3HT as seen in figure S17.

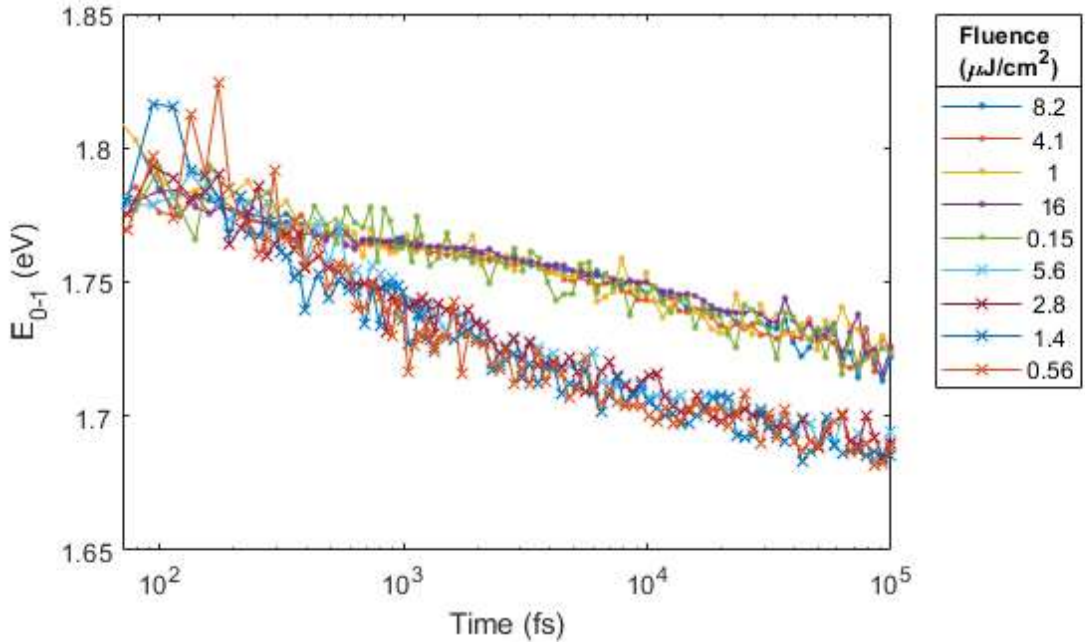

**Figure S17: Fluence dependence of red-shifting.** Comparison of position of $E_{0-1}(t)$ peak extracted from SE in fs-TA. No power dependence is observed for neither the NFs (dots) nor rr-P3HT (crosses).

5.3 Exciton-exciton annihilation (EEA) in fs-TA

Next we turn to exciton-exciton annihilation (EEA); a key process that shortens lifetimes at higher fluences. When analysing annihilation, the first necessary step is to calculate the initial exciton density in film of drop-casted nanofibers. Since the probe used is much smaller than pump, the full-width half-maximum ($FWHM$) of a Gaussian fit is used to find the area of the pump excitation via $A = \pi \left(\frac{FWHM}{2}\right)^2$, as per the standard approximation. The number of excitons, $N_{tot}$ is taken by measuring the pump spectrum and then computing:

$$N_{tot} = \frac{P_0}{f} \int \frac{1}{E} Pump(E)\left(1 - 10^{-OD(E)}\right) dE \qquad (5.5)$$

where $f$ is the pump repetition rate; $Pump(E)$ is the normalized pump spectrum, with an average total power of $P_0$; and $OD(E)$ is the optical density. Next, we must consider the total volume of nanofibers available for the excitons to be distributed through. This is done by first considering a 'calibration' solution containing a known mass concentration $c_{cal}$ with a measured absorption of $OD_{cal}$. Clearly, we must have

$$\frac{OD}{OD_{cal}} = \frac{c}{c_{cal}} \qquad (5.6)$$

So, the mass $m$ contained in the area $A$ is simply

$$m = c_{cal} A l \times \frac{OD}{OD_{cal}} \tag{5.7}$$

where $l$ is the thickness of the cuvette used to the calibration. The total mass is then simply related to the available volume $V$ by the packing density of the nanofiber:

$$V = \frac{m N_A s w h}{3 M_{unimer}} \tag{5.8}$$

where $N_A$ is Avogadro's number; $s, w, h$ are the unit-cell spacing, total width, and total height of the nanofiber respectively; the factor of 3 accounts for the fact that nanofiber is 3 layers high; and $M_{unimer}$ is the molar mass of the unimer. In accordance with previous results we take $s = 0.38$ nm, $w = 12.8$ nm, and $h = 4.5$ nm. Put together, the initial 3-dimensional density of excitons (interrogated by the probe), $n(t = 0)$, is therefore

$$n(0) = \frac{3 M_{unimer} P_0 OD_{cal}}{f c_{cal} A l. OD. N_A s w h} \int \frac{1}{E} Pump(E)\left(1 - 10^{-OD(E)}\right) dE \tag{5.9}$$

We must also consider the effect of anisotropy which will affect the measured $\Delta T/T$ over time. The anisotropy $r$ is found via

$$r(t, \lambda) = \frac{DTT(t, \lambda)_\parallel - DTT(t, \lambda)_\perp}{DTT(t, \lambda)_\parallel + 2\, DTT(t, \lambda)_\perp} \tag{5.10}$$

where of course $DTT \equiv \Delta T/T$.

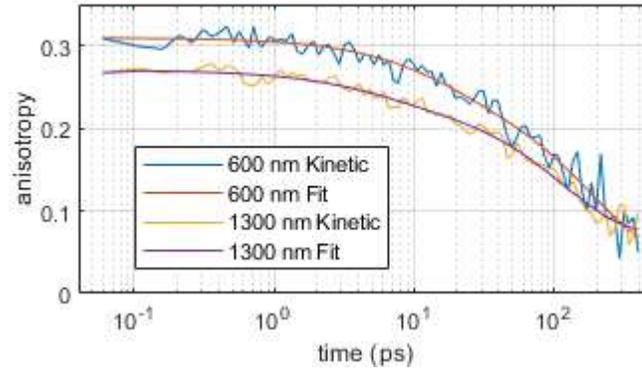

**Fig. S18: Anisotropy kinetics of nanofiber film.** Anisotropy kinetics of nanofibers at $\Delta T/T$ at two spectral regions: 590-610 nm (centred at 600 nm) and 1200-1400 nm (centred at 1300 nm), with tri-exponential fits.

Figure S18 gives the calculated $r(t)$, which is arbitrarily fitted with three exponentials offset by a constant to yield a function $r_{fit}(t)$. The anisotropy decay will be a complex convolution of the hopping of excitons between adjacent nanofibers in the films, and diffusion of excitons along nanofibers that are not perfectly straight. Since this behaviour will be complex, the *exact nature* of the anisotropy decay was not considered further.

Overall, the density of excitons is then given by:

$$n(t) = n(0) \times \frac{3\ln(DTT(t)+1)}{2r_{fit}(t)+1} \times \frac{2r_{fit}(0)+1}{3\ln(DTT(0)+1)} \qquad (5.11)$$

The correction factor of $\frac{3}{2r_{fit}(t)+1}$ converts a TA kinetic taken in the parallel pump/probe configuration (employed to maximize signal-to-noise) to a kinetic that corresponds to a "magic angle" configuration, where the effect of anisotropy decay is removed from the TA kinetic. The $\frac{\Delta T}{T}$ kinetic was taken from the broad peak 1200-1400 nm as this was determined to best represent the decay of singlet excitons.

The exciton dynamics in the presence of annihilation obey the following:

$$\frac{dn(t)}{dt} = -\frac{n(t)}{\tau} - \frac{1}{2}\gamma(t)n(t)^2 \qquad (5.12)$$

with $\gamma(t)$ as the annihilation parameter. In rr-P3HT the diffusion is known to be primarily one-dimensional, in which case $\gamma(t)$ is given by:(7, 27, 54)

$$\gamma(t) = \pi R^2 \sqrt{\frac{8D}{\pi t}} \qquad (5.13)$$

where $D$ is of course the diffusion constant, and $R$ is critical annihilation radius. Since there exists no robust method for determining $R$, for the sake of consistency we will proceed by taking the same value previously used by Tamai *et al.* (*27*) for the crystalline phase of rr-P3HT, where we conservatively take $R = 3.4$ nm. Later, we will give a range of $D$'s for a range of possible $R$ values, but for now it is most instructive to proceed with a single value of $R$. The solution to eq. (11) is

$$n(t) = \frac{n(0)e^{-\frac{t}{\tau}}}{1 + n(0)\pi R^2\sqrt{2D\tau}\,\text{Erf}\left\{\sqrt{\frac{t}{\tau}}\right\}} \qquad (5.14)$$

where Erf{...} is the error function. Eq. 14 is globally fitted to the calculated $n(t)$'s at several different fluences, with each fit sharing the same value of $\tau$ and $D$. In the global fitting procedure, $n(0)$ for each fit is allowed to be slightly optimized (although it is highly constrained by the overall magnitude of each $n(t)$). The minimizer function used is "fminsearch" in a Matlab environment where the sum of the square of the *fractional* residuals between the experiment and theory is minimized in order to not overweigh larger fluences and early times. We also note that there is a certain flexibility in how the global analysis can be implemented; the full measured time range of 400 ps can be used or a more limited time range of 200 ps, and many fluences can be used or a more limited set. For the purposes of full transparency, we show the global fits applied over several different ranges of the data set in fig. S19.

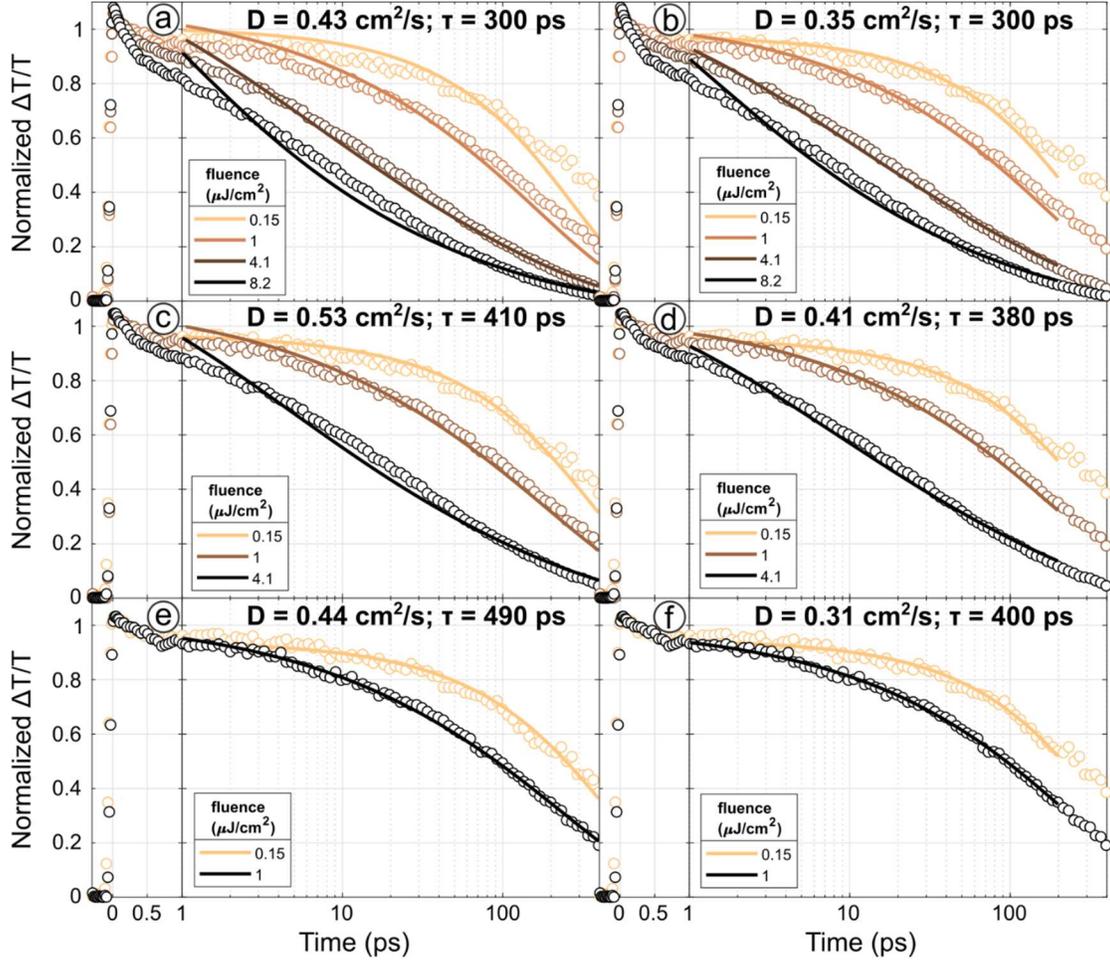

**Fig. S19: Examples of global fits across different ranges of the data set with insets of extracted $D$ and $\tau$ values in each case for $R = 3.4$ nm. (a), (b):** global fits to four fluences from 1-400 ps and 1-200 ps respectively. **(c), (d):** global fits to three lowest fluences from 1-400 ps and 1-200 ps respectively. **(e), (f):** global fits to two lowest fluences from 1-400 ps and 1-200 ps respectively. Considerable variation is observed, with $D$ ranging from 0.3 to 0.5 cm²/s, and $\tau$ ranging from 300 ps to 500 ps.

Figure S19 clearly shows that there is considerable variation in the values for $D$ and $\tau$. Because of this, we elected to give $D$ and $\tau$ in the main text to an accuracy of just one significant figure with uncertainties determined by the range which $D$ and $\tau$ vary over several different implementations of the global fitting (0.3-0.5 cm²/s and 300-500 ps respectively). This results in $D = 0.4 \pm 0.1$ cm²/s and $\tau = 400 \pm 100$ ps.

The reason that the first 1 ps is not included in any of the global fits is that (even at the very lowest fluences) it contains a very small dip on this ultrafast timescale. This is possibly the result of an imperfect anisotropy correction, a coherent artefact, or perhaps a non-equilibrium process that occurs on ultrafast timescales due to excitation above the bandgap (the pump is centred at 540 nm vs a ~600 nm band gap). In any case, this very small dip is not well captured by eq. 5.14, and so we exclude it from the analysis.

We also note that the global fits appear to perform more poorly for the case of high fluences, particularly at early times (1-10 ps), with eq. 5.14 underestimating the amount of decrease in the first 1 ps as we move to higher and higher fluences where multi-exciton phenomena become more and more dominant. This is probably due to introduction of additional decay mechanisms besides diffusion-mediated annihilation at higher fluences. For example, it is likely some excitons undergo 'static' annihilation at the highest fluences, where excitons are created at such a high concentration that a certain proportion of excitons are created already close enough to annihilate one another without even having to move.

As mentioned previously, the value of $R$ is difficult to determine, and values used in the literature vary from around 1-4 nm (*27, 45, 54*). A previously used value of $R = 3.4$ nm extracted by Tamai *et al.* we believe is physically realistic however, and in the main text we conservatively take a range of possible $R$ values from 3 to 4 nm. Taking the value of $D = 0.4$ cm$^2$/s extracted previously using a specific value of $R = 3.4$ nm, we then accordingly find that the range of 3 to 4 nm gives a range of $D = 0.2$ to $1.0$ cm$^2$/s as given in the main text. We do note though that lower values of $R$ are sometimes used – e.g. 1.8 nm – which would result in $D$ values as high as 5 for fs-TA (*45*). We therefore emphasize that TAM – which is a direct measurement of exciton transport – provides the most accurate $D_{NF}$ value.

5.4 Excited-state exciton-phonon couplings in fs-TA

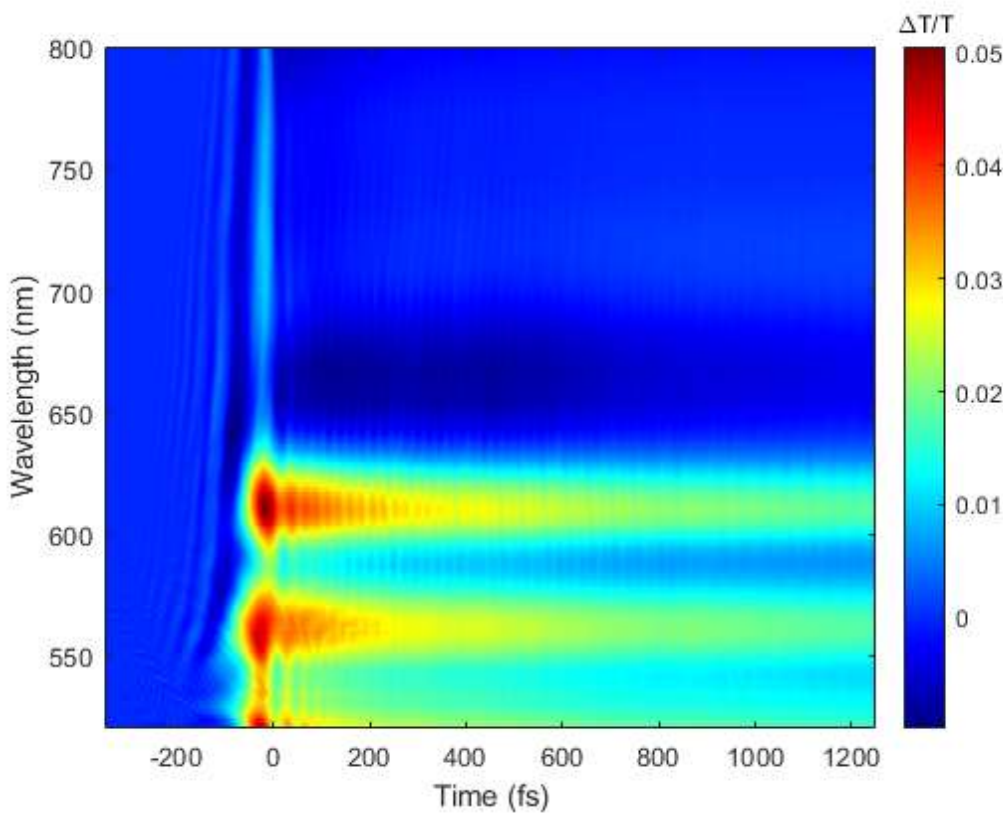

**Fig. S20: fs-TA of NF film using temporally compressed pump pulse.** The pump is centred at 550 nm with a 17 fs duration (measured using frequency-resolved optical gating). The bandwidth is limited to 40 nm to reduce scattering at spectral regions of interest, and a 4 fs step-size is used. The map is the average of 4 individual jitter-corrected sweeps. Pronounced oscillations in $\Delta T/T$ are observed over time at all spectral regions (primarily with a period of 23 fs), which are due to the excited-state vibrational coherences generated by the compressed pump pulse.

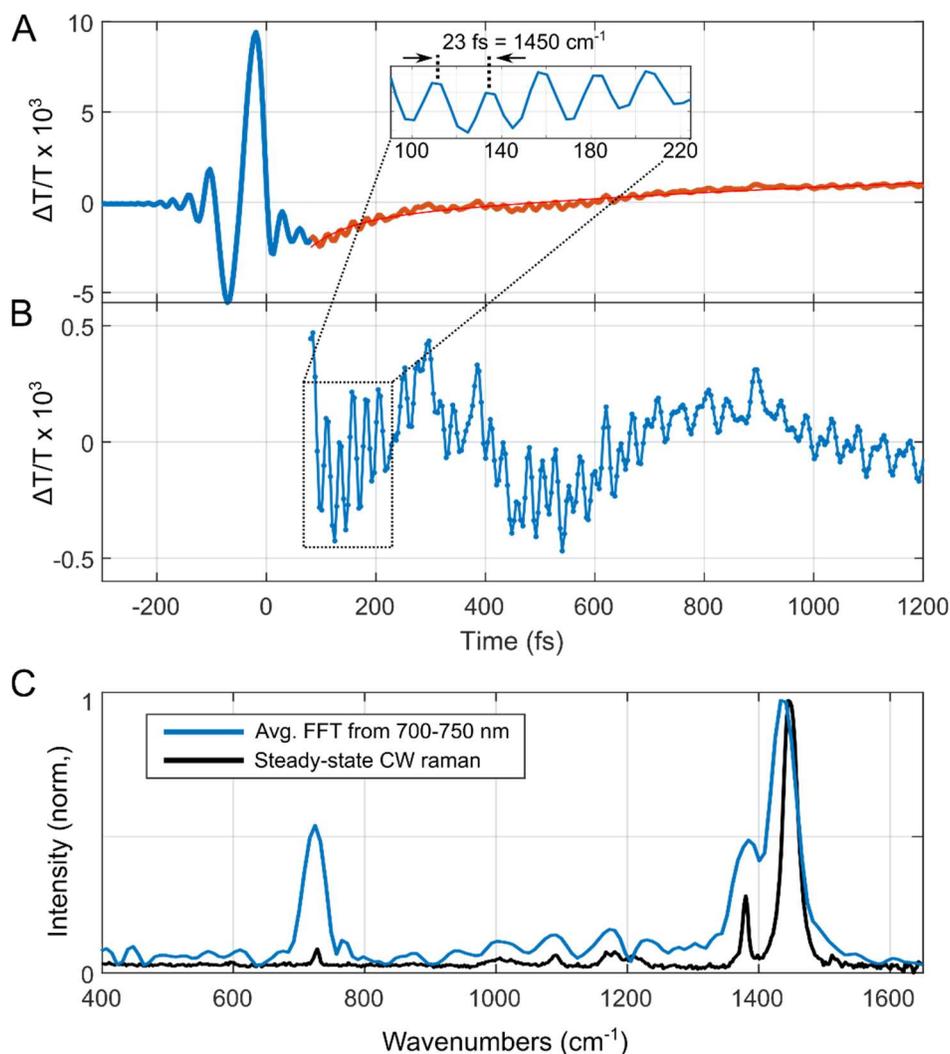

**Fig. S21: Analysis of vibrational coherences in stimulated-emission region (700-750 nm).**
**(A)**: TA kinetic from 700-750 nm taken from data in Fig. S20, with data used to extract coherences highlighted in orange with red bi-exponential fit to model the electronic decay. **(B)**: residual between data and fit in **(A)**, with inset displaying the clear, strong oscillations that occur with a period of 23 fs, which is equivalent to mode with a frequency of 1450 cm$^{-1}$. **(C)** Impulsive Raman spectrum (blue), which gives the Raman modes coupled to the excited state, along with the steady-state continuous-wave (CW) Raman spectrum (black; 532 nm excitation) for comparison. Good agreement between the two is observed, with strong coupling of the 1450 cm$^{-1}$ mode to the excited state, validating our inclusion of this mode in the theoretical simulations. Assignment of the Raman modes to the *excited-state* is justified on the basis of our analysis of the 700-750 nm spectral region, which corresponds to the stimulated-emission of the excited-state. The impulsive Raman spectrum is generated by fitting each of individual $\Delta T/T$ kinetics at each wavelength from 700-750 nm with bi-exponentials, taking the fast-Fourier transform with zero-padding and a Kaiser-Bessel window function (alpha = 2) at each wavelength, and averaging together each of the Raman spectra from 700-750 nm. Low wavenumbers are not considered in this case due to limited time window of 1.2 ps employed which is advantageous for resolving high-frequency modes, but can yield Fourier artefacts for low wavenumbers.

# 6 Ultrafast transient-grating photoluminescence spectroscopy (TRPL)

<u>6.1 TRPL spectra of NFs and controls</u>

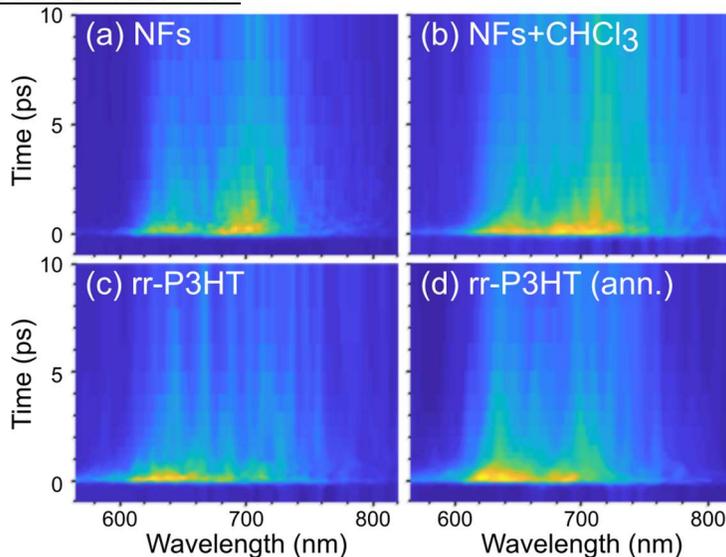

**Fig. S22: TRPL spectra of NFs and three controls samples.** Fluence is 16 $\mu J/cm^2$ (ann. is shorthand for annealed at 170°C for 40 mins).

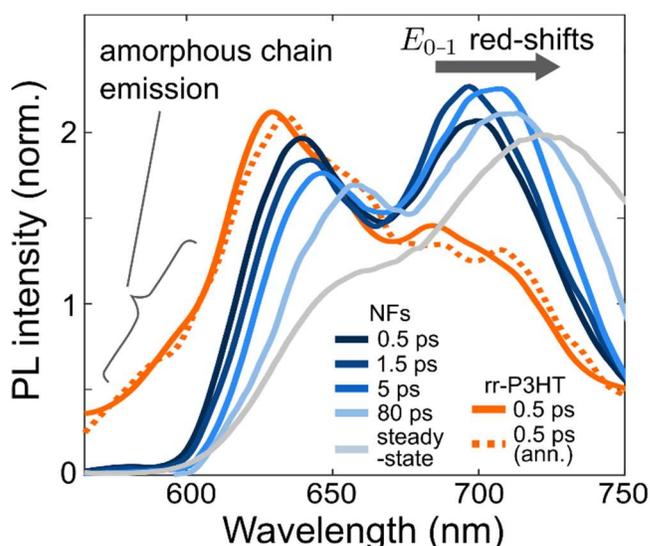

**Fig. S23: Red-shifting in TRPL.** Demonstration of red-shifting in TRPL (normalized by area) collected using a 515 nm excitation at 16 $\mu J/cm^2$, with a Savitzky-Golay filter applied to aid visual interpretation. The steady-state emission is overlaid for easy comparison. An appreciable amount of high-energy emission at very early times is observed for rr-P3HT due to hot carriers excited in the amorphous region of rr-P3HT (*48*, *55*), with annealing at 170°C for 40 mins (ann.) having little effect. The NFs show no such signature, signalling that the NFs contain little or no amorphous domains, and are exclusively crystalline.

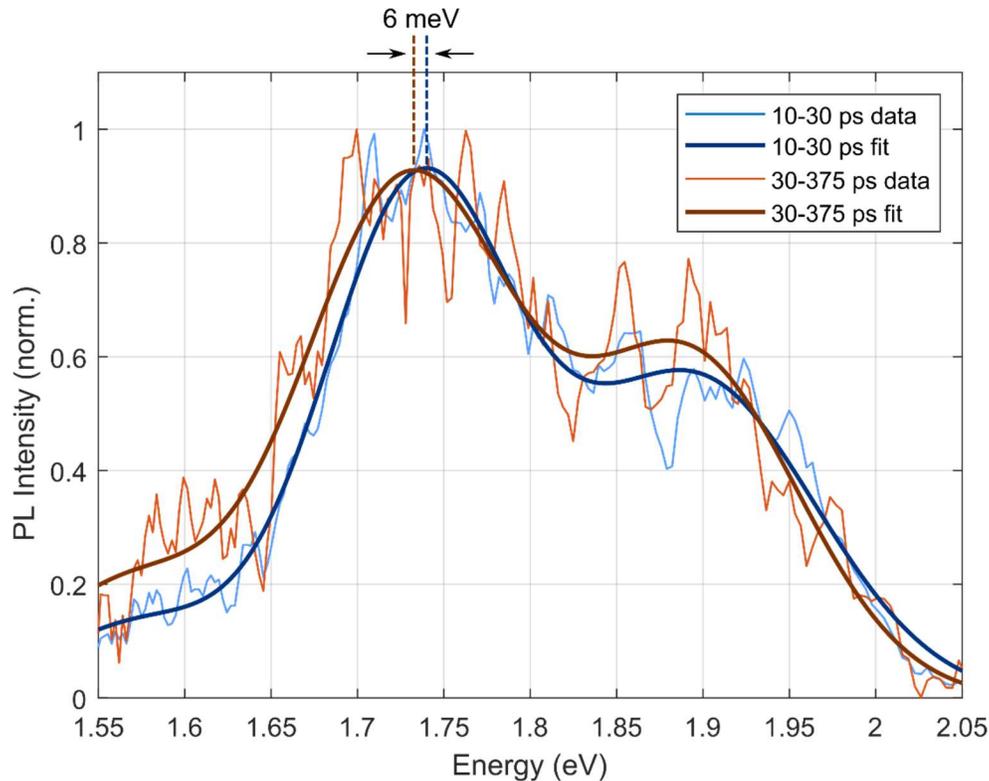

**Fig. S24: Red-shifting compared at two different time perods.** Comparison of unfiltered TRPL spectra of NF film averaged from 10-30 ps and 30-375 ps, collected using a 515 nm excitation at 16 $\mu J/cm^2$. Multi-Gaussian fits are shown in each case. The $E_{0-1}$ peak is found to shift only 6 meV between these two time ranges.

6.2 EEA kinetics in TRPL

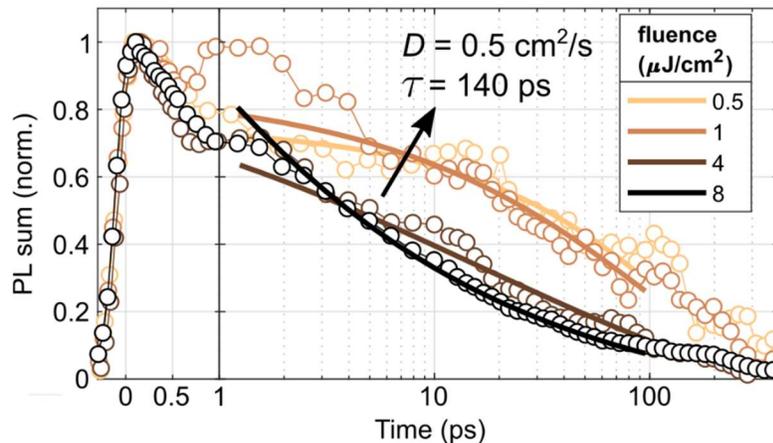

**Fig. S25: Annihilation in TRPL.** Annihilation kinetics of TRPL of NFs from 565 nm to 850 nm, with global fits applied to 1-100 ps. The global fits yield at diffusion constant of $D = 0.5$ $cm^2$/s and a lifetime of $\tau = 140$ ps for the case of $R = 3.4$ nm.

The annihilation kinetics in TRPL are fitted by first converting the PL decays into exciton decays in the same manner as was done for fs-TA i.e. using eq. 5.9. The same eq. 5.14 was used for the global fits. Since TRPL is a difficult experiment we observe substantially more noise, which reduces the fidelity of the fits. However, applying *global* fits over a range of fluences is an effective way to extract $D$ with a reasonable degree of accuracy. To estimate uncertainties, we use the same method as used previously for fs-TA by constructing the global fits over different ranges of the data set. This gives $D = 0.5 \pm 0.2$ cm²/s and $\tau_{fSPL} = 140 \pm 50$ ps for a specific value of $R = 3.4$ nm. As explained previously, for the main text this $D$ value is converted into a range of $D = 0.2$ to $1.0$ cm²/s for $R$ ranging from 3 to 4 nm. The $D$ value (at $R = 3.4$ nm) agrees well with the one found using fs-TA, while the $\tau_{fSPL}$ is reduced somewhat from the value from fs-TA of $400 \pm 100$ ps. This can be understood through two important factors: firstly, it is challenging in TRPL to reach low fluences and to measure to long times because of S/N, and so this prevents the generation of a data set that contains measurements with essentially no EEA. Thus, $\tau$ is not well constrained (as it is in TRPL), and there is tendency for global fits to therefore underestimate $\tau$ in TRPL. Furthermore, there is a possibility of air contamination since the samples used in TRPL experienced several international flights around the world (this was not the case for our other experiments). Although all possible precautions were taken, it is possible air contamination was not fully suppressed. As an illustration of the effect of air contamination, fig. S26 shows how a deliberate exposure to air for one day drastically lowers lifetimes.

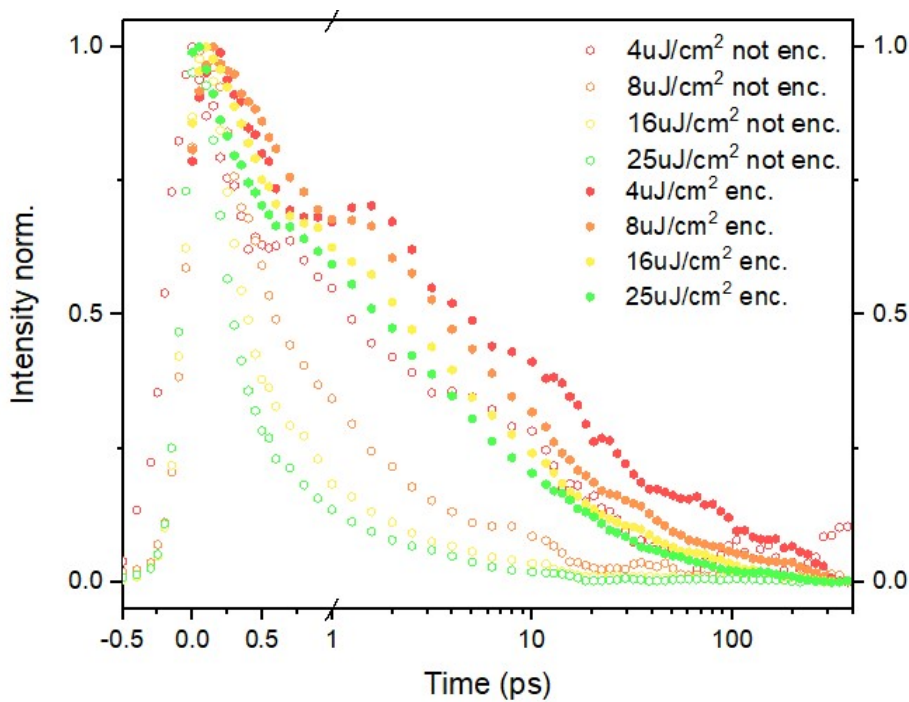

**Fig. S26: Effect on PL kinetics of air exposure.** The intensity of the sum of the PL from the NFs over different fluences. "enc." specifies that the sample is encapsulated in a $N_2$ atmosphere, and "not enc." specifies that there was NO encapsulation and therefore the sample was deliberately exposed to air for ~1 day. We clearly see that air contamination results in much faster decays at the same fluence.

# 7 Transient-absorption microscopy (TAM)

## 7.1 Overview

In a transient-absorption experiment, the $\Delta T/T$ signal is proportional to the pump-induced carrier population (in this case excitons, see ref. (*56*)). In conventional ensemble-level transient-absorption spectroscopy the physical shape of the carrier distribution is not probed and only the kinetics and spectral information of the $\Delta T/T$ signal are recorded. However, in the transient-absorption microscopy (TAM) experiment, we record $\Delta T/T$ as a function of spatial position on a thin film at different pump-probe time delays – see ref. (*25*). This means that we extract out the exciton population at each spatial point at each pump-probe delay. By "expansion of the $\Delta T/T$ distribution" we mean the shape of the $\Delta T/T$ map (shown in Fig. 2B in the main text) expands in size as visualized in the inset in Fig. 2A in the main text. Since $\Delta T/T$ is proportional to the exciton population, it follows mathematically that if the $\Delta T/T$ distribution expands spatially, so too must the exciton distribution. This means that excitons are moving in real space and we can track this transport via monitoring the spatial profile of the $\Delta T/T$ signal.

## 7.2 Fitting of data

As outlined in the main text, in transient-absorption microscopy, images of the $\Delta T/T$ spatial distribution at 600 nm are taken at a succession of pump-probe delays. Each image is then fitted with the following 2D rotated elliptical Gaussian, $f(x,y)$, to effectively measure the width of the distribution:

$$f(x,y) = A \, \text{Exp}\left\{-\frac{(x\cos\alpha + y\sin\alpha + x_0')^2}{2\theta_{x'}} - \frac{(-x\sin\alpha + y\cos\alpha + y_0')^2}{2\theta_{y'}}\right\} \qquad (7.1)$$

where $A$ is the amplitude, $\alpha$ is rotation angle of the Gaussian, $\theta_{x'}$ and $\theta_{y'}$ are the variances along rotated $x'$ and $y'$ axes respectively, and $x_0'$ and $y_0'$ are centre points in the rotated basis. The average variance, $\theta$, of the distribution is then taken as:

$$\theta(t) = \frac{\theta_{x'}(t) + \theta_{y'}(t)}{2} \qquad (7.2)$$

To interpret $\theta(t)$ we first consider our system as one that exhibits isotropic diffusion with a single, exponentially-decay species (excitons). In this is case our system will obey

$$\frac{\partial n(t,x,y)}{\partial t} = D\nabla^2 n(t,x,y) - \frac{1}{\tau} n(t,x,y) \qquad (7.3)$$

If the initial population, $n(0,x,y)$ is a (normalized) Gaussian i.e.

$$n(0,x,y) = \frac{1}{2\pi\sigma(0)^2} \text{Exp}\left\{-\frac{x^2 + y^2}{2\sigma(0)^2}\right\} \qquad (7.4)$$

then the corresponding solution is

$$n(t,x,y) = \frac{1}{2\pi(2Dt + \sigma(0)^2)} \text{Exp}\left\{-kt - \frac{x^2 + y^2}{2(2Dt + \sigma(0)^2)}\right\} \qquad (7.5)$$

where $k \equiv 1/\tau$. Clearly, $n(t, x, y)$ is still a Gaussian at every point in time. The variance of $n(t, x, y)$ is exactly

$$\theta(t) = 2Dt + \sigma(0)^2 \tag{7.6}$$

From this relation the commonly known linear relationship between the mean-square displacement and variance falls out i.e.

$$MSD \equiv \theta(t) - \theta(0) = 2Dt \tag{7.7}$$

As a final note, the standard error, $S$, in the $MSD$ is given through the standard error propagation as

$$S_{MSD}(t) = \sqrt{\left(S_{\theta_x}(t)\right)^2 + \left(S_{\theta_y}(t)\right)^2 + \left(S_{\theta_x}(0)\right)^2 + \left(S_{\theta_y}(0)\right)^2} \tag{7.8}$$

where $S_{\theta_i}(t)$ is the standard error in $\theta_i(t)$ found when fitting $f(x, y)$ to the data.

7.3 Full results from TAM and identification of region independent of fluence effects

An important factor in the TAM experiments is that moderately-high fluences are needed to obtain satisfactory S/N. This results in a substantial amount of EEA. Fig. 8.1a demonstrates how (in concordance with results in fs-TA), faster $\Delta T/T$ decays are observed at higher fluences for the NFs, and this results in larger $\theta$'s being fitted and anomalous expansions in the $MSD$. In particular, higher fluences results in rapid increases in the $MSD$ in the first 0-2 ps. In the next 2-10 ps, smaller increases in the $MSD$ are observed with higher fluences, and from 10-30 ps, it appears that larger fluences do not have any effect on the gradient. The separation into three time regions is of course purely arbitrary, but does serve as a valuable method for interpreting the data. This $MSD$-fluence dependence is interpreted as an effect of EEA, since the exciton density at the centre of the distribution is larger than that at the edges. Because EEA scales via $n^2$ this will result in fast decays at the centre, leading to an artificial broadening in the distribution. Accordingly, when Gaussians are fitted to the distribution, larger $\theta$'s are returned when a substantial amount of EEA at the centre has occurred. Most importantly, EEA is most dominant at very early times due to a variety of factors: the $n^2$ dependence, the radiative decay of excitons, static annihilation (which becomes important at very high fluences), the diffusion of excitons, and the $t^{-1/2}$ dependence of the annihilation parameter. This explains why most of the fluence-related effects on the $MSD$ are seen at very early times.

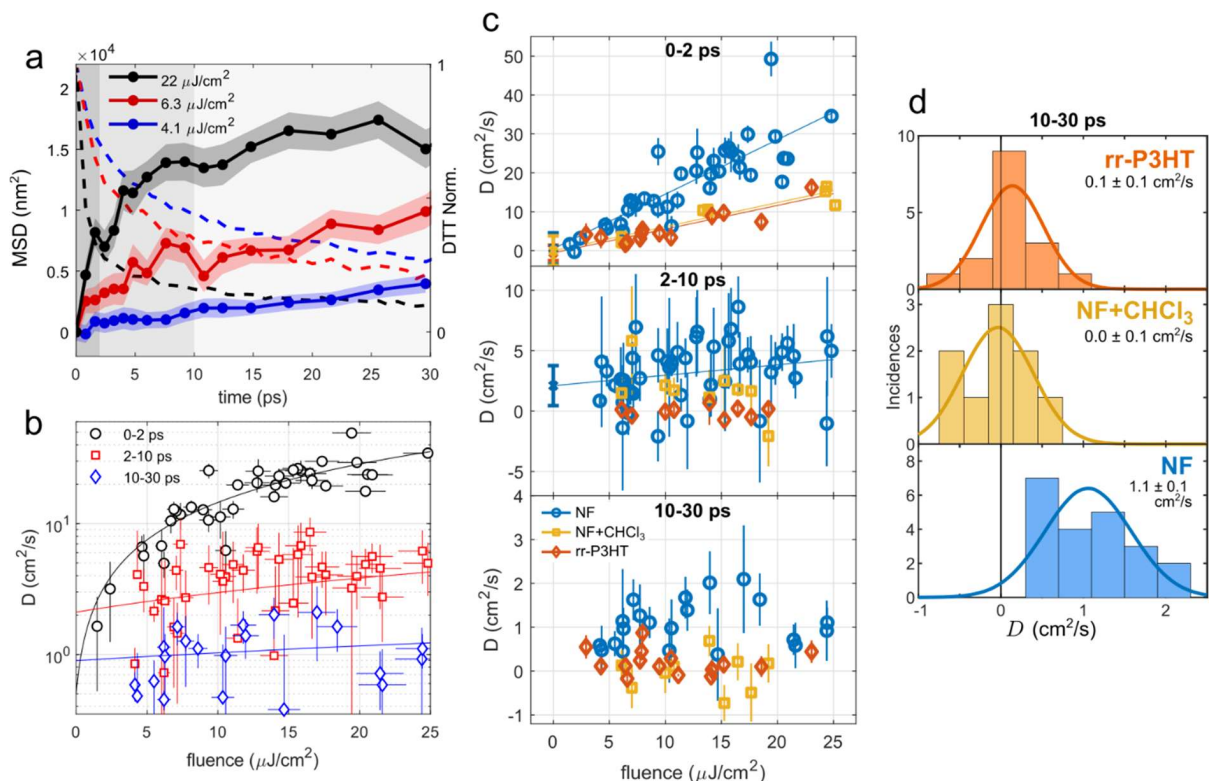

**Fig. S27: TAM data of nanofiber films and controls.** (**a**) MSD and $\Delta T/T$ kinetics (normalized) at centre for several representative experiments on the NFs at different fluences (all fluences indicated are those at the FWHM of the pump distribution). Note the rapid expansion at higher fluences around 0-2 ps, the slightly elevated expansion from 2-10 ps, and the equivalent linear expansion from 10 ps onwards. The $\Delta T/T$ kinetics illustrate how the anomalous expansion is related to the elevated levels of EEA in the NFs at higher fluences. Uncertainties are indicated by shaded regions corresponding to each trace. (**b**) extracted $D$ values for experiments on the NFs found by linear fitting to the regions 0-2 ps, 2-10 ps, and 10-30 ps. Linear fits are displayed in each case. We see a very strong dependence of $D$ with fluence at 0-2 ps, less so from 2-10 ps, and negligible dependence at 10-30 ps. (**c**) extracted values of $D$ in each time period for the NFs and the two control samples: NFs treated with chloroform to disrupt the long-range order, and rr-P3HT. Clear differences are observed between the two controls in comparison to the NFs. In the cases where there is clear dependence, linear fits along with extrapolations to the limit of zero fluence are shown. The horizontal error bars are excluded in this plot to avoid visual clutter, but in general there are 5% uncertainties in the fluence. (**d**) Histogram of the $D$ values in (c) in the time period of 10-30 ps (a histogram is used since there is no clear dependence of $D$ for any of the three samples in this time period). The average values of $D$ in each case are inset. Uncertainties are calculated by dividing by the square root of the number of samples as per usual.

We can visualize the *MSD*-fluence dependence by linearly fitting each of the three regions to extract out quasi-diffusion constants. Strictly-speaking, these are not 'diffusion constants' (in the first two regions) since the broadening of the distribution is due to EEA effects, but we emphasize that such fits provide a very useful way of understanding the data. As shown in fig. 8.1b, the extracted $D$ values for the NFs in the first 2 ps show a very strong dependence on fluence. This dependence is so extreme that we cannot reliably extrapolate $D$ back to zero fluence, (this zero limit would be the 'true' diffusion constant which is due to the actual diffusion of excitons, not apparent broadening due to EEA). However, the dependence of $D$ with fluence is much reduced from 2-10 ps, and we appear to be able to reliably extrapolate back to zero fluence to a find a positive value of $D$, albeit with large error bounds. Most importantly, the dependence of $D$ with fluence from 10-30 ps is negligible. Hence, in this regime we treat the extracted $D$ values as true diffusion constants, and construct a histogram of the values as shown in fig. 8.1d. The mean value here is $D_{NF} = 1.1 \pm 0.1$ cm$^2$/s, which is the value quoted in the main text. This analysis is further supported by the fact that at the lowest fluence used – 4 $\mu$J/cm$^2$/s (see S27a and the main text) – the $MSD(t)$ profile appears to be entirely linear, which can only be the case if diffusion is the only mechanism contributing to increases in the $MSD$.

We also performed the same set of experiments on two control samples – rr-P3HT and chloroform-treated-NFs – to test for the possible existence of systematic experimental artefacts/confirm the large values of $D_{NF}$. Fig. 7.1c,d summarizes the results using the same analysis as described above. On the whole, we see clear differences between the behaviour of the NFs in comparison to the controls, which demonstrates that $D_{NF}$ is *not* an experimental artefact, and is in fact related to the pristine morphology and crystal packing of the NFs. In particular, a substantial dependence of the extracted $D$ values with fluence is observed in the controls in 0-2 ps. This dependence is lower than that of the NFs. We interpret the dependence for the controls as a result of mostly static annihilation, where excitons are created at high densities such that they do not have to move to annihilate each other, and complete this action on ultrafast timescales. Since the dependence for the NFs is larger in 0-2 ps, we suggest that some of the dependence in the NFs is a result of diffusion-mediated annihilation. In the next time period from 2-10 ps, the $D$ values for rr-P3HT show no dependence, and are low and indistinguishable from zero. The $D$ values for the NFs+CHCl$_3$ appear to be non-zero but are substantially lower than that of the NFs. Finally, from 10-30 ps, *both* controls now exhibit $D$ values that are low and indistinguishable from zero, irrespective of the fluence. The histogram in fig. 8.1d also displays the $D$'s from 10-30 ps, along with the mean values; for rr-P3HT we return $D_{rr-P3} = 0.1 \pm 0.1$ cm$^2$/s and for NFs+CHCl$_3$, $D_{NF+CH_3} = 0.0 \pm 0.1$ cm$^2$/s. Such values agree with what is expected; for rr-P3HT, the literature reports values on the range of 10$^{-2}$-10$^{-3}$ cm$^2$/s, which are currently below the resolution of TAM (*26, 27, 32*). For the NFs+CHCl$_3$, a low value (< 0.1) is also be expected if the superior performance of the NFs is due to their superior energetic and long-range positional order.

Note that we interpret $D_{NF}$ to be the inter-chain diffusion of excitons. This is because inter-fiber exciton diffusion is expected to be much weaker due to the much weaker couplings between chains of different nanofibers. Intra-chain diffusion will also not contribute significantly to $D_{NF}$ since the chains are short, and any movement of energy within the chain is likely to reach an equilibrium very quickly.

As a final comment, we return to the subject of polarons as mentioned for the fs-TA experiments. The fs-TA data shows that the yield of polarons is negligible due to photo-excitation at 540 nm, and that the yield due to EEA is also small. This is to be expected in a neat material that contains very little energetic inhomogeneity (which is typically required for the generation of charges in organic semiconductors). However, in the interests of maximum transparency we entertain the question: "is the expansion shown in fig. 8.1b a result of the (unbiased) diffusion of charges that are produced from EEA?" We believe that such a question is immediately disqualified by the fact that such a scenario would result in a super-linear increase in $D$ with fluence since the generation of charges from EEA is expected to follow a $n^2$ dependence, and this super-linear increase is clearly not observed. Furthermore, $D = 1.1 \pm 0.1$ cm$^2$/s would imply – via the Einstein mobility relationship – a spectacular mobility of 40 cm$^2$ V$^{-1}$ s$^{-1}$, which is greatly in excess of the highest values demonstrated for rr-P3HT (about 0.1 cm$^2$ V$^{-1}$ s$^{-1}$) (*57*). In reality, the mobility would have to be much larger than 40 cm$^2$ V$^{-1}$ s$^{-1}$ since polarons are greatly in the minority compared to excitons. Overall, such a hypothetical scenario is not physically realistic (our simulations show that exciton transport is enhanced by long-range couplings, which are *not* present for charge transport), and this scenario is not supported by the data, so we exclude any effects of charges.

In the interests of clarity, we have also shown the diffusion constants found from the 10-30 ps region below in Table S1:

| Diffusion constants extracted from 10-30 ps (cm$^2$/s) | | |
|---|---|---|
| rr-P3HT (control) | NF+CHCl$_3$ (control) | Pristine NFs |
| 0.128957973 | 0.55127955 | 1.137305845 |
| -0.377106481 | 0.120228296 | 1.108334088 |
| -0.039589945 | 0.112963663 | 0.58530877 |
| -0.727113329 | 0.870945184 | 0.48110756 |
| 0.687886473 | -0.121419229 | 1.264309515 |
| 0.104065301 | -0.171277858 | 0.967102571 |
| 0.213937588 | 0.240987232 | 0.624592363 |
| 0.17604012 | -0.089607067 | 0.979668419 |
| -0.489043526 | 0.289575451 | 0.451163981 |
| | 0.448873751 | 1.631515068 |
| | 0.108243085 | 0.469639139 |
| | 0.151239743 | 1.62702159 |
| | 0.033891094 | 0.922049751 |
| | 0.101792248 | 2.096270209 |
| | 0.443124508 | 0.378738274 |
| | -0.860626995 | 1.674881739 |
| | | 1.390305297 |
| | | 2.014752031 |
| | | 0.584698896 |
| | | 0.715176966 |
| | | 1.106510283 |

**Table S1:** Diffusion constants extracted from 10-30 ps (cm$^2$/s) as shown in the histogram in Fig. S27d. The numbers are not rounded to prevent rounding errors when computing the mean and standard deviations.

# 8 Theoretical modelling

We consider stacks of P3HT chains as one-dimensional arrays of equally spaced $N$ sites (with an open boundary condition), where each P3HT chain is coarse-grained as a single site.

## 8.1. Steady-state optical absorption and emission spectra

We follow ref. (*58*) to model the steady-state optical properties of the P3HT one-dimensional stacks. In brief, we use the two-particle approximation to solve the Frenkel-Holstein model, accounting for a single effective vibrational mode with energy $\hbar\omega_1 = 1450$ cm$^{-1}$:

$$H = \sum_k \varepsilon_k |k\rangle\langle k| + \sum_{k,l;l>k} V_{k,l}(|k\rangle\langle l| + |l\rangle\langle k|) + \omega_1 \sum_k b_k^+ b_k + \omega_1 \lambda_1 \sum_k (b_k^+ + b_k)|k\rangle\langle k| + \omega_1 \lambda_1^2 \tag{8.1}$$

$\varepsilon_k$ is the gas phase $S_0 \to S_1$ transition energy (including the gas-to-aggregate transition energy shift due to nonresonant second-order dispersion forces). $V_{k,l}$ is the exciton coupling between chains $k$ and $l$, $|k\rangle$ is a pure electronic state of the aggregate with chain $k$ excited to $S_1$ while all others remain in their ground-state $S_0$, $b_k^+$ ($b_k$) is the creation (annihilation) operator for the $S_0$ vibrational quanta on chain $k$, $\lambda_1 = F_1^{1/2}$ is the exciton-phonon coupling constant related to the Huang-Rhys factor $F_1$.

The Coulomb excitonic couplings are obtained from the atom-centered INDO/SCI transition densities calculated for the lowest electronic excitation in a 30-mer polythiophene chain:

$$V_{k,l} = \frac{1}{4\pi\varepsilon_0} \sum_{a\epsilon k, b\epsilon l} \frac{\rho_{ka}\,\rho_{lb}}{r_{kl,ab}} \tag{8.2}$$

where $\rho_{ka}$ is the transition density on atom $a$ of chain $k$ and $r_{kl,ab}$ is the distance between atoms $a$ in chain $k$ and atom $b$ in chain $l$. The high-frequency mode corresponds to the dominant carbon-carbon breathing vibration in P3HT and its energy $\hbar\omega_1$ is adjusted to reproduce the main vibronic progression measured in solution. Eq. (8.1) is diagonalized to obtain the eigenstates $|\psi_j\rangle$ and associated eigenenergies $\varepsilon_j$. The dimensionless polarized absorption spectrum is then calculated as:

$$A(\omega) = \langle (NW_a(0)\mu^2)^{-1} \sum_j |\langle G|\widehat{M}|\psi_j\rangle|^2 W_a(\omega - \varepsilon_j)\rangle \tag{8.3}$$

$\langle ... \rangle$ denotes average over the energy distribution. $\widehat{M}$ is the transition dipole moment operator of the aggregate, which allows one-photon excitation from the aggregate ground state, $|G\rangle$, to the $j$-th eigenstate, $|\psi_j\rangle$, and $W_a(\omega)$ is the absorption line shape function. The emission spectrum takes the form:

$$S(\omega) = \langle (W_e(0))^{-1} \sum_{v_t, t=0,1,2,...} I(v_t)(1 - v_t\omega_0/\omega_{em})^3 W_e(\omega - \omega_{em} + v_t\omega_0)\rangle \tag{8.4}$$

$\omega_{em}$ is the frequency of the lowest-energy emitting state, that is we assume rapid population relaxation down to the lowest state $|\psi^{(em)}\rangle$ prior to emission. $W_e(\omega)$ if the emission line shape function and $I(v_t)$ the dimensionless $0 - v_t$ line strength given by:

$$I(v_t) = \mu^{-2} < \sum_{\{v_n\}}' \left|\langle\psi^{(em)}|\hat{M}\prod_n|g_n,v_n\rangle\right|^2 \tag{8.5}$$

Here, $\prod_n|g_n, v_n\rangle$ is a terminal state in the emission event in which the polymer chain $n$ is in the electronic ground state with $v_n$ vibrational quanta in the ground state potential. The prime on the summation in eq. (8.5) indicates the constraint $\sum_n v_n = v_t$.

The lineshape functions in absorption and emission, $W_a(\omega)$ and $W_e(\omega)$ respectively, include both homogeneous and inhomogeneous contributions. The (dynamic) homogeneous contribution is likely associated with conformational degrees of freedom along single P3HT chains while the (static) inhomogeneous contribution might have various origins, such as the presence of structural kinks or chemical defects. We are interested in ensemble averaged properties ($\langle ... \rangle$), hence the optical spectra are averaged over the disorder taken to be uncorrelated and Gaussian. The total linewidth is $\sigma^2 = \sigma_{ho}^2 + \sigma_{inhom}^2$. The inhomogeneous disorder contributes to the Stokes shift between absorption and emission, as the latter only occurs from thermalized excitons (i.e., at the bottom of the exciton Density of States) while both dynamic and static contributions affect the shape of the spectra. A reasonable fit to experiment is obtained when considering $\sigma_{hom} = 0.06$ eV and $\sigma_{in} = 0.036$ eV in conjunction with $F_1 = 0.9$, see Figure S28. We note, however, that the calculations overestimate the energy of the emitting excitons (underestimate the Stoke shift), in particular at long times, which has been associated with differences in the torsion potentials in the ground vs excited state of the polymer chains (ref. (*58*)).

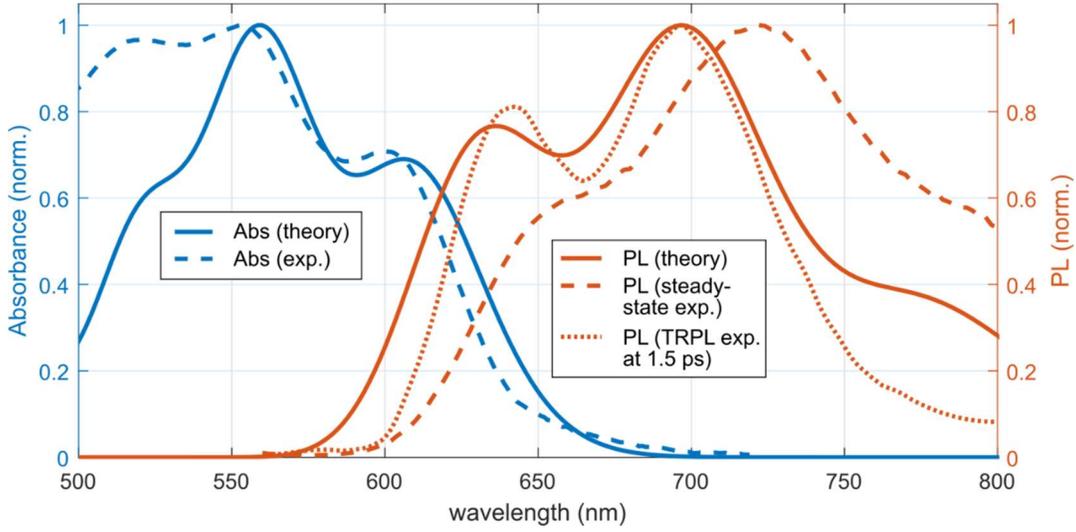

**Fig. S28: Comparison of absorption and emission calculated via the Frenkel-Holstein model with experiment.** The time-resolved PL spectrum from TRPL at 1.5 ps is also given to highlight how the red-shifting of the PL over time (which is a complex convolution of this ability for the excitons to diffuse over time, and the way disorder is distributed spatially) complicates the comparison between the theoretically-calculated PL and the steady-state PL.

8.2. Non-adiabatic exciton transport simulations

As mentioned in the main text, we resort to the crossing-corrected subspace surface hopping algorithm (*39–41*), a state-of-the-art variant of the traditional fewest switch surface hopping (FSSH) algorithm developed by Tully (*42*) to simulate the non-adiabatic molecular dynamics of exciton diffusion in NFs. Within this mixed quantum-classical technique, nuclear degrees of freedom are treated classically while the electronic degrees of freedom are probed quantum-mechanically. The nuclear dynamics of the system is described by an ensemble of non-interacting trajectories, each time-evolving under the influence of a single, 'active' adiabatic potential energy surface (PES) maintaining the Newtonian dynamics, while the electronic wavefunction is propagated along the trajectory via the time-dependent Schrödinger equation (TDSE). Non-adiabatic transitions between different adiabatic PESs are stochastically incorporated maintaining the internal consistency, *i.e.*, the fraction of trajectories on each PES should conform to the corresponding quantum population obtained by TDSE.

To proceed, we separate out the total Hamiltonian in Eq. (8.1) as:

$$H = H_e + H_n \tag{8.6}$$

where the electronic Hamiltonian is represented by

$$H_e = \sum_k \left(\varepsilon_k + \alpha_1 x_{k,1} + \alpha_2 x_{k,2}\right)|k\rangle\langle k| + \sum_{k,l;l>k} V_{k,l}(|k\rangle\langle l| + |l\rangle\langle k|) \tag{8.7}$$

while the nuclear Hamiltonian is given by

$$H_n = \sum_k \frac{1}{2}\left[m_{k,1}\dot{x}_{k,1}^2 + K_1 x_{k,1}^2 + m_{k,2}\dot{x}_{k,2}^2 + K_2 x_{k,2}^2\right]. \tag{8.8}$$

The electronic Hamiltonian is expressed in the local diabatic basis $\{|k\rangle\}$, where $|k\rangle$ represents that the exciton is localized on the $k$-th polymer chain (as described in the previous subsection). In the current study, we consider two effective intramolecular, classical, vibrational degrees of freedom – one high frequency mode $(x_{k,1})$, the carbon-carbon stretching mode at energy $\hbar\omega_1$ and one low-frequency mode $(x_{k,2})$, which is the source of spectral broadening (homogeneous contribution). Thus, the model Hamiltonian in Eqs. 8.4-8.6 is formally equivalent to that in Eq. 8.1, yet the high-frequency vibration is treated quantum-mechanically in Eq. 8.1 and classically in Eq. 8.4-8.6. The corresponding force constants, effective masses and velocities are denoted by $K_{1(2)}$, $m_{k,1(2)}$ and $\dot{x}_{k,1(2)}$, respectively. $\alpha_1$ and $\alpha_2$ are the corresponding local exciton-phonon couplings. $V_{k,l}$ and $\varepsilon_k$ have the same meaning as above. Diagonalizing the electronic Hamiltonian for a particular set of $\{x_{k,1}\} \equiv \mathbf{x_1}$ and $\{x_{k,2}\} \equiv \mathbf{x_2}$, we obtain the adiabatic states $\{|\psi_i(\mathbf{x_1}(t), \mathbf{x_2}(t))\rangle = \sum_k p_{k,i}|k\rangle\}$ with adiabatic energies $\{E_i(t)\}$, which can be used as the basis to represent the excitonic wavepacket $|\Psi(t)\rangle = \sum_i w_i(t)|\psi_i(\mathbf{x_1}(t), \mathbf{x_2}(t))\rangle$. The TDSE in the adiabatic basis is given by

$$\dot{w}_i(t) = \left[\frac{w_i(t)E_i(\mathbf{x_1}(t), \mathbf{x_2}(t))}{\imath\hbar}\right] - \sum_{j\neq i} w_j(t)\left[\sum_k (\dot{x}_{k,1} \cdot d_{ij}^{k,1} + \dot{x}_{k,2} \cdot d_{ij}^{k,2})\right] \tag{8.9}$$

with $d_{ij}^{k,1(2)} = \langle \psi_i(\mathbf{x_1}(t), \mathbf{x_2}(t))|\partial\psi_j(\mathbf{x_1}(t), \mathbf{x_2}(t))/\partial x_{k,1(2)}\rangle$ being the non-adiabatic couplings, which are obtained using the Hellmann−Feynman theorem (analytical expressions are similar to Eq. 6 in ref. (59)). The analytical expression of the excitonic ground state energy $E_1 = \langle \psi_1|H|\psi_1\rangle$ is

$$E_1 = E_1^{(0)} + E_1^{(1)} + E_1^{(2)}, \tag{8.10}$$

$$E_1^{(0)} = \sum_k \varepsilon_k p_{k,1}^2 + \sum_{k,l;l>k} 2V_{kl}\, p_{k,1} p_{l,1}, \tag{8.11}$$

$$E_1^{(1)} = \sum_k \alpha_1 x_{k,1} p_{k,1}^2 + \sum_k \frac{K_1}{2} x_{k,1}^2, \tag{8.12}$$

$$E_1^{(2)} = \sum_k \alpha_2 x_{k,2} p_{k,1}^2 + \sum_k \frac{K_2}{2} x_{k,2}^2. \tag{8.13}$$

The initialization procedure comprises of relaxation of the exciton to the bottom of the energy band and definition of initial nuclear coordinates and velocities for individual trajectories. The central chain in the stack is initially excited to $S_1$ with $x_{N/2,1(2)} = -\alpha_{1(2)}/K_{1(2)}$, $x_{k\neq N/2,1(2)} = 0$, following by iterative update of $\mathbf{x_1}$ and $\mathbf{x_2}$ till $E_1$ is minimized (40, 41). The energy minimum can be easily found analytically with $x_{k,1(2)} = -\alpha_{1(2)} p_{k,1}^2/K_{1(2)}$. The optimized structure is stored. For each surface hopping trajectory at temperature $T$, the initial $x_{k,1(2)}$ is set following the Boltzmann distribution with variance $k_B T/K_{1(2)}$ around the optimized $x_{k,1(2)}^{(opt)}$, while the initial velocities $\dot{x}_{k,1(2)}$ of site $k$ are also chosen randomly from a Boltzmann distribution with variance $k_B T/m_{k,1(2)}$, $k_B$ being the Boltzmann constant (59, 60).

The step-by-step outline of the surface hopping algorithm is as follows:

(1) **Nuclear dynamics:** The nuclear equation of motion is modelled following the Langevin equation (59):

$$m_{k,1(2)} \ddot{x}_{k,1(2)} = -\frac{\partial \langle \psi_a | H | \psi_a \rangle}{\partial x_{k,1(2)}} - \gamma m_{k,1(2)} \dot{x}_{k,1(2)} + \xi_{1(2)} \tag{8.14}$$

where '$a$' represents the active adiabatic state, $\gamma$ is the friction coefficient characterizing the system-bath interaction and $\xi_{1(2)}$ is the Markovian Gaussian random force with standard deviation $(2\gamma m_{k,1(2)} k_B T/dt)^{1/2}$, $dt$ being the time step size. Standard fourth-order Runge-Kutta method is employed to solve the Langevin equations (for more details, see ref. (59)).

(2) **Wavepacket propagation:** Since solution of the TDSE is representation independent, time propagation of the wavepacket is more suitable in the diabatic basis as the non-adiabatic couplings are not involved and convergence of the solution can be achieved with larger time step size(61, 62). We adopt the widely used locally diabatic basis proposed by Granucci et al., which has shown fast time step size convergence (63, 64). In this algorithm, the adiabatic states at each time $t$ ($\{|\psi_i(x_1(t), x_2(t))\rangle\}$) are utilized as basis states to expand the wavefunction at time $\tau$ within the time interval $[t, t+dt]$

$$|\Psi(\tau)\rangle = |\psi_i(x_1(t), x_2(t))\rangle. \tag{8.15}$$

The wavefunction can be simply propagated through

$$w_i^{LD}(\tau) = w_i(t) \exp\left(\frac{E_i(t)(\tau - t)}{\iota \hbar}\right), \tag{8.16}$$

since the Hamiltonian in the adiabatic representation is diagonal. Then the wavefunction can be converted back to the adiabatic representation (a more suitable basis for calculation of the hopping probabilities) by $w_i(\tau) = \sum_j w_j^{LD} U_{ji}^*$, where $U_{ji} = \langle \psi_j(x_1(t), x_2(t)) | \psi_i(x_1(\tau), x_2(\tau)) \rangle$.

(3) **Subspace construction:** We construct subspaces of adiabatic states within the time interval $[t, t+dt]$.(40) $M$ states with highest quantum populations at time $t$ ($P_i(t) = |w_i(t)|^2$) are chosen and the current active state $a$ is included manually. Namely, if it is not within the subspace, the selected state with the smallest population is replaced by the active state. Then we calculate the absolute overlap of all adiabatic states $j$ at time $t+dt$ with these selected states at time $t$, $S_j = \sum_{k=1}^{M} |\langle \psi_k(t) | \psi_j(t+dt) \rangle|$ and $M$ adiabatic states with the highest overlaps are chosen. These two sets of adiabatic states at time $t$ and $t+dt$ constitute the subspaces for surface hopping. The states within each subspace are re-indexed in ascending order of energy and the corresponding physical quantities are updated from the values in full space.

(4) **Surface hopping and velocity adjustment:** The crossing-corrected algorithm (40, 41) is employed to carry out the stochastic hopping since large systems suffer from the severe trivial crossing problem. First, the hopping probabilities from the active state '$a$' to all other adiabatic states '$i$' in the subspace are calculated within the traditional FSSH approach (*42*):

$$g_i = -\frac{2dt\,\Re\left[w_a^*(t)w_i(t)\sum_k\left(\dot{x}_{k,1}\cdot d_{ai}^{k,1} + \dot{x}_{k,2}\cdot d_{ai}^{k,2}\right)\right]}{|w_a(t)|^2}. \tag{8.17}$$

Next, we calculate the overlap between the active state $a$ at time $t$ and every adiabatic state in the subspace at time $t + dt$ and find the state with the maximum overlap ($j$). If $j \neq a$, a trivial crossing with the active state is encountered and therefore, the hopping probabilities are self-consistently corrected (ref. 9) as $g_j = [|w_a(t)|^2 - |w_a(t+dt)|^2]/|w_a(t)|^2 - \sum_{i\neq j} g_i$. If $g_j < 0$, then they are reset to zero.

A uniform random number $\zeta \in [0,1]$ is generated and hopping to a state $b$ is assigned if the condition $\sum_{i=1}^{b-1} g_i < \zeta \leq \sum_{i=1}^{b} g_i$ is satisfied. If $b = j$, then the hopping is immediately accepted, and the new active state index becomes $j$. Otherwise, we find the adiabatic state $k$ in the subspace at time $t + dt$, which has the maximum overlap with state $b$ at time $t$. If the nuclear velocities can be adjusted along the non-adiabatic coupling vector, $k$ becomes the new active state index; otherwise the system remains on $j$-th adiabatic surface. The adjusted nuclear velocities after the hop from active state $a$ to adiabatic state $b \neq j$ are given by

$$\dot{x}'_{k,1(2)} = \dot{x}_{k,1(2)} + d_{ab}^{k,1(2)}\frac{A}{B}\left[\sqrt{1 + \frac{2(E_a - E_b)B}{A^2}} - 1\right], \tag{8.18}$$

where $A = \sum_k\left(\dot{x}_{k,1}d_{ab}^{k,1} + \dot{x}_{k,2}d_{ab}^{k,2}\right)$ and $B = \sum_k\left[\left(d_{ab}^{k,1}\right)^2/m_{k,1} + \left(d_{ab}^{k,2}\right)^2/m_{k,2}\right]$.

(5) **Decoherence correction:** Decoherence correction is an essential part of surface hopping algorithm to achieve higher accuracy. In our study, we have considered the energy-based decoherence algorithm proposed by Truhlar and coworkers (*65*) and further simplified by Granucci and coworkers (*66*). At each time step, the decoherence time for an adiabatic state $i \neq a$ is determined by

$$\tau_i = \frac{\hbar\left(1 + \frac{C}{E_{kin}}\right)}{|E_i - E_a|}, \tag{8.19}$$

where $E_{kin}$ is the total nuclear kinetic energy and $C$ is set to 0.1 Hartree (*65, 66*). The wavefunction coefficients are then adjusted by

$$w'_i = \begin{cases} w_i \exp\left(-\frac{dt}{\tau_i}\right) & i \neq a \\ \dfrac{w_i(1 - \sum_{k\neq a}|w'_k|^2)}{|w_i|} & i = a \end{cases} \tag{8.20}$$

and converted back to the local diabatic basis for post-calculation analysis.

(6) **Convergence:** Steps 1-5 are repeated until the predetermined criterion is achieved.

The mean-squared displacement (MSD) of the exciton is calculated from the time-propagated wavepacket by

$$MSD(t) = \frac{1}{N_{traj}} \sum_{i=1}^{N_{traj}} \langle \psi_a^{(i)}(t) | r^2 | \psi_a^{(i)}(t) \rangle - \left[ \frac{1}{N_{traj}} \sum_{i=1}^{N_{traj}} \langle \psi_a^{(i)}(t) | r | \psi_a^{(i)}(t) \rangle \right]^2, \quad (8.21)$$

where $N_{traj}$ is the number of independent surface hopping trajectories and $\psi_a^{(i)}$ is the active adiabatic state at time $t$ for the $i$-th trajectory. The matrix elements can be computed with $\langle k | r^2 | l \rangle = \delta_{kl} k^2 L^2$ and $\langle k | r | l \rangle = \delta_{kl} k L$, $L$ being the spacing between nearest neighbors and $\{|k\rangle\}$ are the aforementioned diabatic basis states. To obtain a smooth profile of $MSD(t)$, we have considered 10,000 trajectories with time step size ($dt$) of 0.01 fs and the subspace size $M = 10$. Linear evolution of $MSD(t)$ signifies that an equilibrium diffusion regime is attained, and the exciton diffusion coefficient ($D$) is calculated by

$$D = \frac{1}{2} \lim_{t \to \infty} \left[ \frac{d(MSD(t))}{dt} \right]. \quad (8.22)$$

The time-dependent inverse participation ratio (IPR) is a proxy for the exciton delocalization length along the polymer stack and is calculated by

$$IPR(t) = \frac{1}{N_{traj}} \sum_{i=1}^{N_{traj}} \frac{1}{\sum_k \langle k | \psi_a^{(i)}(t) \rangle^4}. \quad (8.23)$$

As mentioned before, we have considered a high frequency mode of $\hbar\omega_1 = 1450$ cm$^{-1}$, and a low frequency mode of $\hbar\omega_2 = 50$ cm$^{-1}$. Corresponding vibrational force constants are $K_1 = 9.30$ mdyne/Å and $K_2 = 0.011$ mdyne/Å, the effective masses are $m_{k,1} = m_{k,2} = 7.50$ amu; These parameter values are comparable to the ab initio results obtained for P3HT hexamer (with methyl side chain) by density functional theory (DFT) calculations at the B3LYP/6-31G** level. The Huang-Rhys factor for the high-frequency modes, $F_1 = 0.9$, is used to deduce $\alpha_1 = 24575.0$ cm$^{-1}$Å$^{-1}$ from $\lambda_1 = \hbar\omega_1 F_1 = \alpha_1^2/K_1$. As described above, the spectral shape can be reproduced by considering $\sigma_{hom} = 0.06$ eV and $\sigma_{inhom} = 0.036$ eV. From $\sigma_{hom}$, we can infer $\alpha_2 = 803.5$ cm$^{-1}$ Å$^{-1}$ using, for a classical harmonic oscillator, $\lambda_2 = \hbar\omega_2 F_2 = \alpha_2^2/K_2 \sim \sigma_{hom}^2/k_B T$. $\gamma = 100$ ps$^{-1}$ is taken in accordance to previous surface hopping studies (*40, 41, 59, 61*) while $L = 3.85$ Å is obtained from the crystallographic data.

8.3. Simulated MSD and IPR profiles

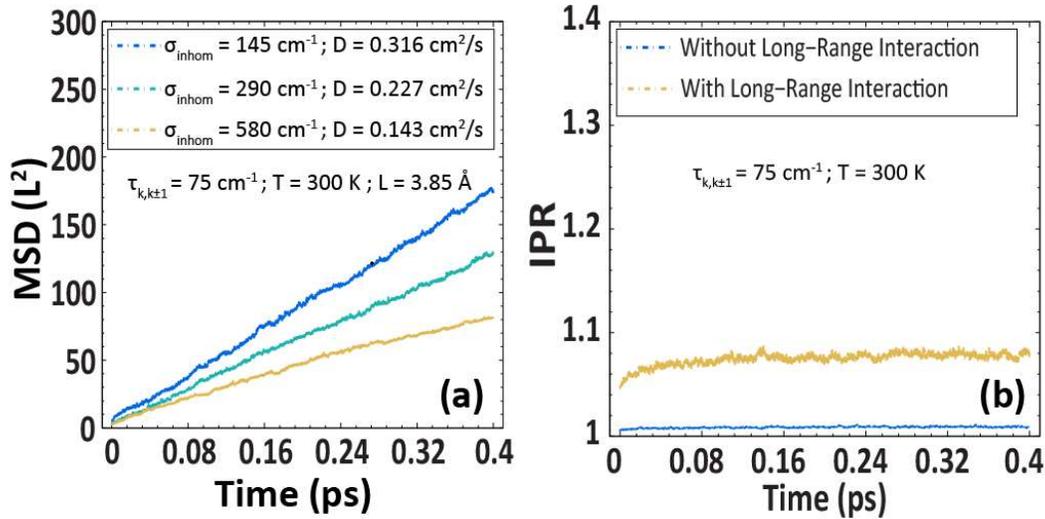

**Fig. S29: MSD profiles from simulations under varying levels of energetic disorder.** Comparison of (**a**) the MSD for different energetic disorder $\sigma_{inhom}$ and (**b**) the average exciton delocalization length in presence and absence of long-range excitonic couplings.

As mentioned in the main text, simulated *MSD* profiles as well as *D* exhibit modest dependence on the inhomogeneous energetic disorder ($\sigma_{inh}$), see fig. S29(a). On increasing $\sigma_{inhom}$ from 290 cm$^{-1}$ (corresponds to NFs) to 580 cm$^{-1}$ (similar to the energetic disorder reported in rr-P3HT system), *D* does not reduce abruptly, suggesting that long-range interactions smooth out, to some extent, energetic inhomogeneity in the system. On the other hand, the time-evolution of the average IPR, both in the presence and absence of long-range interactions, indicates that the excitons remain mostly localized over a single chain in the stack (fig. S29(b)). Due to the transient delocalisation mechanism in presence of long-range interactions, where the system momentarily accesses more delocalized states, the average IPR becomes slightly higher in this case.

We also describe in the main text how the exciton sporadically occupies states with high IPR values. Fig. S30 and S31 provide histograms showing the probabilities that an exciton at any point in the simulation time will attain a particular IPR value. With the inclusion of long-range exciton couplings, the probability that the system attains a delocalized energy state (IPR >= 2) becomes significant (3%), but without long-range couplings the probability is very low (<0.001%). This again clearly points to importance of long-range exciton couplings for allowing for delocalized energy states.

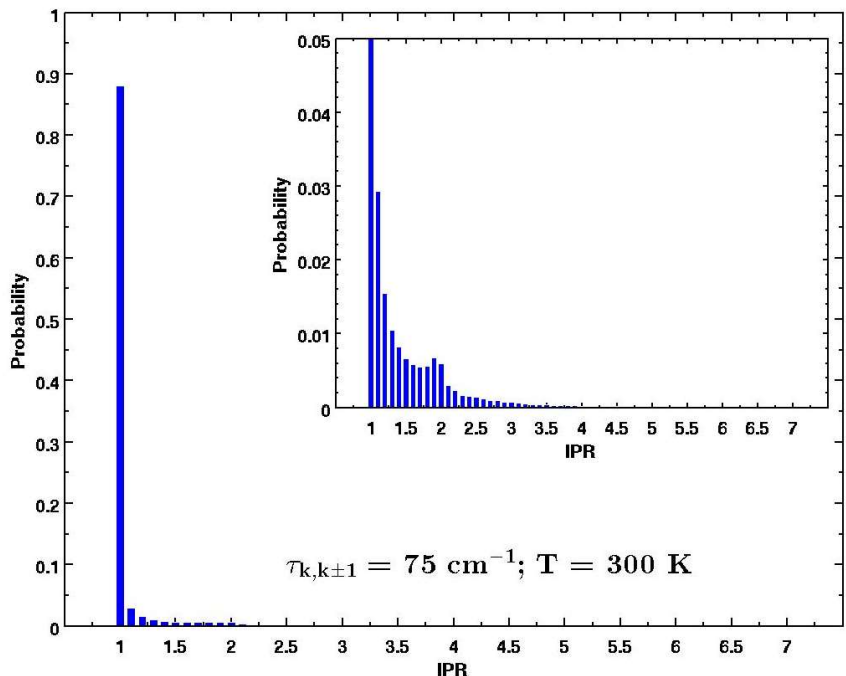

**Fig. S30: Histogram of the time-dependent IPR values of the system with long-range interactions, sampled over ~100 trajectories.** The system attains a delocalized energy state (IPR >=2) ~3% of the simulation time.

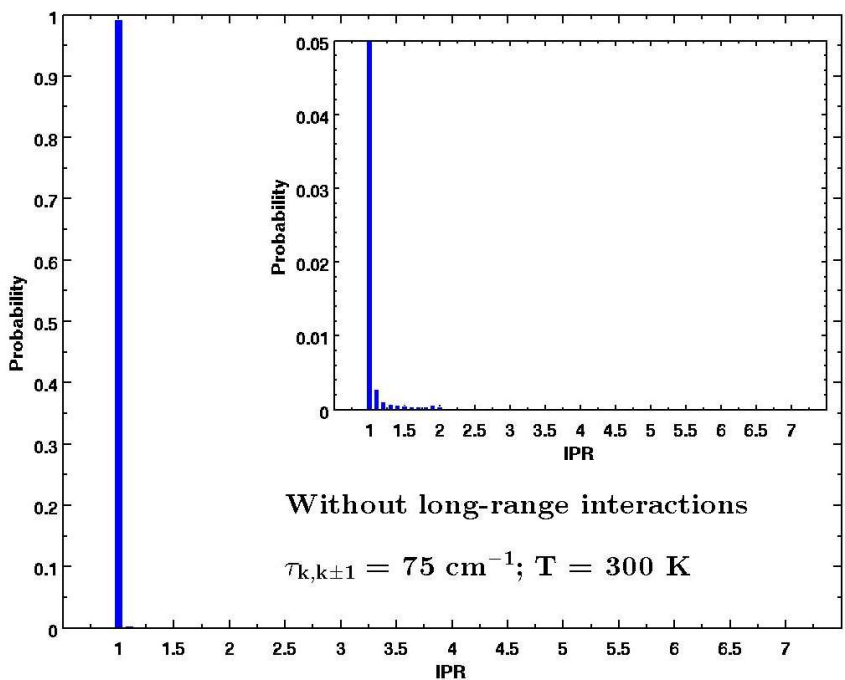

**Fig. S31: Histogram of the time-dependent IPR values of the system without long-range interactions, sampled over ~100 trajectories.** The system attains a delocalized energy state (IPR >=2) <0.001% of the simulation time.

## 9 Förster hopping limit

To calculate the incoherent Förster hopping limit, we first calculate the self Förster radius, $R$, of the NFs via the usual method (*30*):

$$R^6 = \frac{9(\ln 10)\kappa^2 Q_Y}{128\pi^5 N_{AV} \eta^4} J \tag{9.1}$$

where $\kappa$ is the relative orientation between the absorptive and emissive dipoles, and is assumed to be $\kappa = 2$; $Q_Y$ is the quantum yield of the nanofibers, which is measured to be $Q_Y = 0.05$ via an integrating sphere with 405 nm continuous wave excitation, with the $Q_Y$ calculation made via the method described by de Mello *et. al.* (*67*); $\eta$ is the refractive index of the medium, and is taken as $\eta = 2.3$ from ref. (*68*). $J$ is the (self) overlap integral evaluated from nanofibers' absorption and PL – see ref. (*30*) for more details; and $N_{AV}$ is Avogadro's number. We find that the self Förster radius is $R = 1.2$ nm. Using $L_D = R^3/b^2$, (where the interchain spacing, $b$, is 0.38 nm), we estimate a diffusion length of 11 nm. Then, by making use of $L = \sqrt{2D\tau}$ (where $\tau$, the exciton lifetime is taken to be 400 ps from fs-TA measurements), we estimate a value of $D = 2 \times 10^{-3}$ cm²/s for the rate of exciton diffusion predicted from simple Förster hopping. These extracted $D$ and $L_D$ value are similar to previously calculated values that only consider incoherent hopping in P3HT (*28, 31, 38*). Significantly, the $D$ value is three orders of magnitude lower that the experimental value of $D_{NF} = 1.1$ cm²/s. This shows that the simplistic approach described above is a poor description of the NF system, validating the use of our more advanced treatment thoroughly described in section 8 of this document.